\documentclass[fleqn,twocolumn,tighten,apj,twocolappendix]{aastex63} 
	
\usepackage[english]{babel}
\usepackage{blindtext}
\usepackage{CJKutf8}

\usepackage{amsmath,bm,nccmath,textcomp,gensymb}

\usepackage{scalerel}

\usepackage{lipsum,hyperref,multirow,dsfont,balance,enumitem,tabularx,booktabs,dcolumn}

\usepackage[titletoc]{appendix}

\newcommand\apMADGICS{\texttt{apMADGICS}}
\newcommand\korg{\texttt{Korg.jl}}

\shorttitle{\apMADGICS{} RV Precision}
\shortauthors{Saydjari et al.}

\graphicspath{{./}{figures/}}

\begin{document}

\title{Improving Radial Velocities by Marginalizing over Stars and Sky: \\ Achieving 30 m/s RV Precision for APOGEE in the Plate Era}


\correspondingauthor{Andrew K. Saydjari}
\email{andrew.saydjari@princeton.edu}

\author[0000-0002-6561-9002]{Andrew K. Saydjari}
\altaffiliation{Hubble Fellow}
\affiliation{Department of Physics, Harvard University, 17 Oxford St., Cambridge, MA 02138, USA}
\affiliation{Center for Astrophysics | Harvard \& Smithsonian, 60 Garden St., Cambridge, MA 02138, USA}
\affiliation{Department of Astrophysical Sciences, Princeton University,
Princeton, NJ 08544 USA}

\author[0000-0003-2808-275X]{Douglas P. Finkbeiner}
\affiliation{Department of Physics, Harvard University, 17 Oxford St., Cambridge, MA 02138, USA}
\affiliation{Center for Astrophysics | Harvard \& Smithsonian, 60 Garden St., Cambridge, MA 02138, USA}

\author[0000-0001-7339-5136]{Adam J. Wheeler}
\affiliation{Department of Astronomy, Ohio State University, McPherson Laboratory 140 West 18th Avenue, Columbus, OH 43210, USA}

\author[0000-0002-9771-9622]{Jon A. Holtzman}
\affiliation{Department of Astronomy, New Mexico State University, MSC 4500, Box 30001, Las Cruces, NM 88033, USA}

\author[0000-0001-7828-7257]{John C. Wilson}
\affiliation{Astronomy Department, University of Virginia, Charlottesville, VA 22904, USA}

\author[0000-0003-0174-0564]{Andrew R. Casey}
\affiliation{School of Physics \& Astronomy, Monash University}
\affiliation{Center for Computational Astrophysics, Flatiron Institute, 162 5th Avenue, New York, NY 10010, USA}
\affiliation{Centre of Excellence for Astrophysics in Three Dimensions (ASTRO-3D)}

\author[0000-0002-0947-569X]{Sophia Sánchez-Maes}
\affiliation{Center for Astrophysics | Harvard \& Smithsonian, 60 Garden St., Cambridge, MA 02138, USA}

\author[0000-0002-8725-1069]{Joel R. Brownstein}
\affiliation{Department of Physics and Astronomy, University of Utah, 115 S. 1400 E., Salt Lake City, UT 84112, US}

\author[0000-0003-2866-9403]{David W. Hogg}
\affiliation{Center for Computational Astrophysics, Flatiron Institute, 162 5th Avenue, New York, NY 10010, USA}
\affiliation{Center for Cosmology and Particle Physics, Department of Physics, 726 Broadway, Room 1005, New York University, New York, NY 10003, USA}
\affiliation{Max Plank Institute for Astronomy, K\"onigstuhl 17, 69117 Heidelberg, Germany}

\author[0000-0003-1641-6222]{Michael R. Blanton}
\affiliation{Center for Cosmology and Particle Physics, Department of Physics, 726 Broadway, Room 1005, New York University, New York, NY 10003, USA}

\begin{abstract}

The radial velocity catalog from the Apache Point Observatory Galactic Evolution Experiment (APOGEE) is unique in its simultaneously large volume and high precision as a result of its decade-long survey duration, multiplexing (600 fibers), and spectral resolution of $R \sim 22,500$. However, previous data reductions of APOGEE have not fully realized the potential radial velocity (RV) precision of the instrument. Here we present an RV catalog based on a new reduction of all 2.6 million visits of APOGEE DR17 and validate it against improved estimates for the theoretical RV performance. The core ideas of the new reduction are the simultaneous modeling of all components in the spectra, rather than a separate subtraction of point estimates for the sky, and a marginalization over stellar types, rather than a grid search for an optimum. We show that this catalog, when restricted to RVs measured with the same fiber, achieves noise-limited precision down to 30 m/s and delivers well-calibrated uncertainties. We also introduce a general method for calibrating fiber-to-fiber constant RV offsets and demonstrate its importance for high RV precision work in multi-fiber spectrographs. After calibration, we achieve 47 m/s RV precision on the combined catalog with RVs measured with different fibers. This degradation in precision relative to measurements with only a single fiber suggests that refining line spread function models should be a focus in SDSS-V to improve the fiber-unified RV catalog.

\end{abstract}


\section{Introduction} \label{sec:Intro}

Precision measurements of the radial velocities of stars are crucial for detecting exoplanets \citep{Wright_2018_haex}, systems of stellar binaries/multiples \citep{Kounkel_2021_AJ}, stellar streams \citep{Bonaca_2024_arXiv}, and Galactic archaeology \citep{Tremaine_2023_MNRAS}. The first two applications depend predominantly on differential velocity measurements of the same target, which can be performed by cutting-edge instruments at m/s precision \citep{Mayor_2003_Msngr}. The latter two applications rely more on absolute velocities, which involves the application of corrections such as gravitational redshift that involve uncertainties at the 100s of m/s level \citep{AllendePrieto_2013_A&A}. In this work, we will only report ``radial velocity measures,'' as recommended by the IAU, which do not include these corrections \citep{Lindegren_2003_A&A}, but will refer to them colloquially as radial velocities (RVs) throughout.

The cutting-edge precision (differential) radial velocity instruments that have $\leq$ 1 m/s precision like MAROON-X \citep[$R\sim85,000$,][]{Seifahrt_2020_SPIE}, HARPS \citep[$R\sim115,000$,][]{Mayor_2003_Msngr}, and ESPRESSO \citep[$R\sim140,000$,][]{Pepe_2021_A&A} typically only measure a single target at a time. In contrast, most large spectroscopic surveys are fiber multiplexed or slitless, but attain $\sim1$ km/s RV precision like BOSS \citep[$R\sim2,000$, 5 km/s,][]{Almeida_2023_ApJS, kounkelpyxcsao}, DESI \citep[$R\sim3,500$, 1 km/s,][]{Cooper_2023_ApJ}, LAMOST \citep[MRS, $R\sim7,500$, 800 m/s,][]{Zhang_2021_ApJS}, H3 \citep[$R\sim32,000$, 5-10 km/s,][]{Conroy_2019_ApJ}, RAVE \citep[$R\sim7,500$, 1.4 km/s,][]{Steinmetz_2020_AJ}, and Gaia RVS \citep[$R\sim11,500$, 1.3 km/s,][]{Katz_2023_A&A}. 

In this work, we will focus on the Apache Point Observatory Galactic Evolution Experiment (APOGEE) survey \citep{Majewski_2017_AJ}. Targeted surveys with M2FS ($R\sim50,000$) achieved 25-80 m/s RV precision, which is comparable to the precision of APOGEE, but only observed a few hundred stars \citep{Bailey_2016_AJ,Bailey_2018_MNRAS}. The most comparable previous catalog is from GALAH ($R\sim28,000$), which delivered $\sim140$ m/s RV precision and contains $\sim340,000$ stars \citep{Zwitter_2018_MNRAS}. However, an RV catalog for APOGEE DR17  \citep{Abdurro'uf_2022_ApJS} that could achieve its theoretical $\sim30$ m/s RV precision and contains $\sim730,000$ stars pushes further into the high-precision, large-survey frontier on both axes.

Typical pipelines that measure RVs, such as \texttt{doppler} used by the APOGEE data reduction pipeline (DRP), rely on finding peaks in the cross correlation of templates with observed spectra \citep{nidever2021dnidever}. This involves both searches over libraries of theoretical or data-driven stellar templates and proposed radial velocity shifts. These methods can also suffer significantly from residual contamination by imperfectly subtracted sky lines or imperfectly corrected telluric absorption features \citep{Bedell_2019_AJ}. The novel method we present (1) removes the need for grid-based searches over stellar type by marginalizing over stellar type in a single shot and (2) jointly fits the stellar and sky components to reduce biases from tellurics and sky lines. In what follows, we describe the improvements our method brings to the APOGEE RV catalog, increasing the RV precision through software alone.

In Section \ref{sec:Data} we describe the APOGEE spectra used in this work. Section \ref{sec:Methods} introduces the component separation method (Julia package \texttt{apMADGICS.jl}) and RV repeats we use for validation. In Section \ref{sec:RVPrecision} we show the improvement in RV precision obtained by using \apMADGICS{} and discuss the RV catalog calibration in Section \ref{sec:RV_fib_cal}, including a fiber-fiber offset. The calibration of error bars and comparison to theoretical lower bounds on RV as a function of stellar type are shown in Sections \ref{sec:RVcal} and \ref{sec:RVType}, respectively. We demonstrate the benefits of \apMADGICS{} for multi-epoch RV work in Section \ref{sec:RVVary}, describe how to access the data behind all figures in Section \ref{sec:dataavil}, and conclude in Section \ref{sec:conc}, with an eye on upcoming work in SDSS-V. In the appendices we show fiber-to-fiber variations in Gaussian centroids of the LSF (Appendix \ref{sec:LSFCentroid}), apply the ``Q-factor'' analysis from Section \ref{sec:RVType} to an example spectrum and exploration of the definition of S/N (Appendix \ref{sec:QfactorChips}), and comment on the computational details of the ``Q-factor'' computation (Appendix \ref{sec:CompSpeed}).

\section{Data} \label{sec:Data}

In this work we use the results of a general component separation method applied to stellar spectra obtained using the APOGEE spectrographs \citep{Wilson_2012_SPIE,Wilson_2019_PASP}. To obtain full sky coverage, there are two ``copies'' of the instrument, APOGEE-N at Apache Point Observatory (APO) on the Sloan Foundation 2.5 m telescope \citep{Gunn_2006_AJ} and APOGEE-S at Las Campanas Observatory (LCO) on the Irénée du Pont 2.5 m telescope \citep{Bowen_1973_ApOpt}. The APOGEE spectrographs were developed for the Apache Point Observatory Galactic Evolution Experiment survey \citep{Majewski_2017_AJ} spanning SDSS-III \citep{Eisenstein_2011_AJ} and SDSS-IV \citep{Blanton_2017_AJ}, but remain in operation during SDSS-V \citep{Kollmeier_2017_arXiv} as a core part of the Milky Way Mapper survey \citep{Almeida_2023_ApJS}. We will refer to all spectra obtained with either instrument as APOGEE spectra, regardless of their survey origin. For simplicity, we restrict ourselves to spectra released in SDSS data release 17 (DR17), which covers a decade of APOGEE operations \citep{Abdurro'uf_2022_ApJS}.

The APOGEE spectrographs are fed with 300 fibers each, measure near-infrared spectra from $\sim15,000 - 17,000$ \r{A} with resolution $R = \lambda/\Delta\lambda \approx 22,500$, using 3 mercury cadmium telluride (HgCdTe) 2048 $\times$ 2048 pixel detectors, which leave two $\sim50$ \r{A} gaps \citep{Arns_2010_SPIE, Blank_2010_SPIE, Brunner_2010_SPIE, Wilson_2019_PASP}. The detectors are read out non-destructively per pixel by ``sampling-up-the-ramp'' (SUTR) for $\sim 47$ reads per exposure, leading to a 3D data cube. These data are
reduced by the APOGEE ``Data Reduction Pipeline'' (DRP), to which we will often compare \citep{Nidever_2015_AJ}. The reduced data products \apMADGICS{} starts from are the \texttt{ap1D} files which are produced by the DRP per exposure.

Because the APOGEE line spread function (LSF) is not well-sampled by the detectors (FWHM$_{\rm{LSF}} < 3$ pixels), especially for shorter wavelengths at APO, visits for an object consist of exposures that have a ``dither shift'' between them, where the detector is physically shifted by $\sim0.5$ pixels. The detector is typically dithered in an A-B-B-A pattern, with an equal number of exposures at each dither position per visit \citep{Majewski_2017_AJ}. \apMADGICS{} handles this dither combination for ``good'' exposures in a given visit. Each configuration of fibers on the sky is specified by a plate number that corresponds to a physical plate of aluminum, a so-called "plug-plate".  Positioned at the telescope focal plane, fiber connectors are plugged into precisely drilled holes in the plate which corresponding to specific targets on the sky.  While not included here, observations with APOGEE in SDSS-V that will be analyzed in subsequent work will include those that make use of a robotic fiber positioning system which replaced the plug plate system.

\begin{figure*}[t]
\centering
\includegraphics[width=0.97\linewidth]{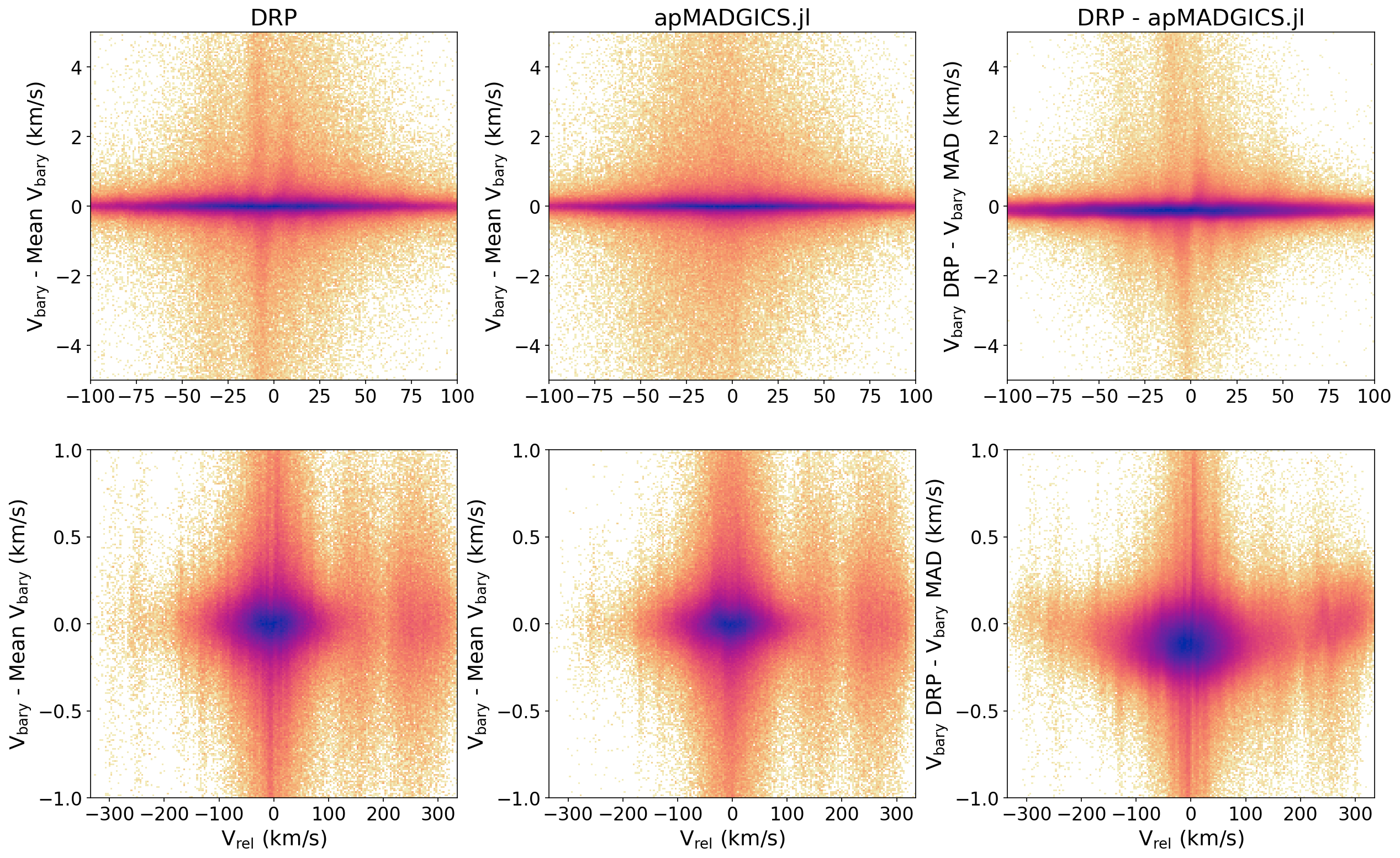}
\caption{Logarithmic 2D histograms of the deviation of the observed barycentric velocity ($v_{\rm{bary}}$) from the average for a given star shown as a function of the sky (observatory) frame velocity ($v_{\rm{rel}}$) for the DRP (left) and \apMADGICS{} (middle). These plots are restricted to stars with more than 5 visits with good RV measurements. The last panel shows the difference between the barycentric velocity from the DRP and \apMADGICS{} for the same visit as a function of the observed (sky frame) velocity. The top and bottom rows show different velocity scales on both axes.
}
\label{fig:RV_DRP_MAD}
\end{figure*}

In our analysis, we apply cuts to restrict to ``good'' RV visits, non NaN \texttt{RV\_verr} for \apMADGICS{} and non NaN \texttt{VHELIO} for the DRP. In \apMADGICS{} \texttt{RV\_verr} is most often NaN when either all of the exposures from the DRP are NaNs or the minimum of the $\chi^2$ surface occurs at a grid edge \citep[see Section \ref{sec:Methods} and][]{APOGEE_pipeline_unpub}. At times, we also restrict to targets well-modeled by ASPCAP by rejecting \texttt{STARFLAG} bits (0,1,3,4,12,13,16-19,29,21-23,25), rejecting \texttt{EXTRATARG} bits (1,2,3), rejecting \texttt{ASPCAPFLAG} bits (0,1,3-5,8,10-13,16-24,26-36,40,41), requiring \texttt{FE\_H\_FLAG} be zero, and requiring $T_{\rm{eff}}$ not NaN. The meaning on these targeting and ASPCAP bits are described on the SDSS-IV website.\footnote{\href{https://www.sdss4.org/dr17/irspec/apogee-bitmasks/}{https://www.sdss4.org/dr17/irspec/apogee-bitmasks/}} We also apply a cut on stellar type requiring $\log(g) > 1.3$ to eliminate the most luminous red giants, which have convective motions and pulsations, and require T$_{\rm{eff}} < 6000$ to eliminate stars with especially broad hydrogen lines that are often not well modeled and can perturb the continuum. The samples are often further outlier-mitigated by requiring $\sigma(v_{\rm{bary}})$ over all visits $< 1$ km/s, where $\sigma(v_{\rm{bary}})$ is estimated by jackknifing as discussed below. Throughout, we will use the median signal-to-noise (S/N) across the whole spectrum to summarize the S/N of a single visit (see Appendix \ref{sec:QfactorChips}).

Everywhere we refer to fibers by ``adjusted fiber index,'' which runs 1-300 at APO and 301-600 at LCO. This numbering scheme matches the physical trace ordering in the read-out direction at each telescope, plus 300 at LCO. This indexing means all data products from 3D to 1D spectra are indexed the same way. However, it is inverted relative to the physical labeling of the fibers, the plugmaps and configuration files, and the historical APOGEE convention. Everywhere we refer to fibers by number in text, we will include the physical FIBERID parenthetically.


\section{Methods} \label{sec:Methods}

\subsection{MADGICS: Component Separation} \label{sec:ComponentSep}

We decompose each visit spectrum using Marginalized Analytic Data-space Gaussian Inference for Component Separation \citep[MADGICS,][]{Saydjari_2023_ApJ}. MADGICS is an algorithm that decomposes an observation, a vector of data $x_{\text{tot}}$, into a linear combination of components $x_k$ (Equation \ref{eq:decomp_set}). To separate a single observation into components requires prior information about each of the components involved in the decomposition. In the context of MADGICS, we express this prior as a pixel-pixel covariance matrix $C_k$ in data space. For APOGEE, this is the covariance of the flux from that component in wavelength bin $i$ and wavelength bin $j$ for a given visit spectrum after \apMADGICS{} has resampled and combined the exposures onto a universal (logarithmic) wavelength axis. In practice, we learn these priors from data, models, or a hybrid of the two, as detailed in \citep{APOGEE_pipeline_unpub}.

Prior information about the mean of each component $\mu_k$ can also be included. Deciding when to impose a prior on the mean and when to set $\mu_k = 0$ is difficult, but extremely important. Choosing to set $\mu_k = 0$ can significantly inflate the prior covariance $C_k$. However, sometimes loose priors are required to describe a component with large variability and heterogeneity, for example, components with multi modal distributions. In practice, we only use a prior on the mean for the per exposure sky continuum and sky lines components \citep{APOGEE_pipeline_unpub}.

For notational convenience, we define the sum of the prior covariances and means over all components in the model (Equation \ref{eq:tot_def}).
\begin{ceqn}
\begin{align} \label{eq:decomp_set}
    x_{\rm{tot}} &= \sum_{k=1}^{N_{\rm{comp}}} x_k \quad \text{where} \quad x_k \sim \mathcal{N}(\mu_k,C_k) \\
    \label{eq:tot_def}
    C_{\rm{tot}} &= \sum_{k=1}^{N_{\rm{comp}}} C_k \quad \quad \mu_{\rm{tot}} = \sum_{k=1}^{N_{\rm{comp}}} \mu_k
\end{align}
\end{ceqn}
\indent To obtain a mean and posterior for the contribution of each component to the overall observation, we define the likelihood to be the product of the Gaussian probability for $\hat{x}_k$ given $\mathcal{N}(\mu_k,C_k)$ for all $k$ components. Then, we impose the key constraint that the components sum exactly to the data and analytically solve to obtain Equations \ref{eq:post_mean}-\ref{eq:post_km}. This flux conservation, that the sum of the components equals the data, is exact to numerical precision for a linear model.
\begin{ceqn}
\begin{align} \label{eq:post_mean}
    \hat{x}_{k} &= C_k C_{\rm{tot}}^{-1}\left(x_{\rm{tot}} -\mu_{\rm{tot}}\right) + \mu_k \\
    \label{eq:post_kk}
    \hat{C}_{kk} &= (I - C_k C_{\rm{tot}}^{-1})C_k \\
    \label{eq:post_km}
    \hat{C}_{km} &= -C_k C_{\rm{tot}}^{-1} C_m
\end{align}
\end{ceqn}
where $I$ is the identity matrix, $\hat{x}_k$ is the posterior mean component, $\hat{C}_{kk}$ is the posterior pixel-pixel covariance of pixels in component $k$, and $\hat{C}_{km}$ is the predicted covariance of a pixel in component $k$ with a pixel in component $m$. From an application stand-point, Equations \ref{eq:post_mean}--\ref{eq:post_km} demonstrate that the component separation depends only on a single matrix inverse $C_{\rm{tot}}^{-1}$, which can be computed quickly by using low-rank approximations and Woodbury updates \citep{max1950inverting}. 

For APOGEE spectra, the relevant components in our model are sky lines, sky continuum, star continuum, star lines, and residuals. For a more detailed description of how the priors are built see \cite{APOGEE_pipeline_unpub}, which fully describes the APOGEE MADGICS pipeline \apMADGICS{} (see Section \ref{sec:dataavil}).

\begin{figure*}[t]
\centering
\includegraphics[width=\linewidth]{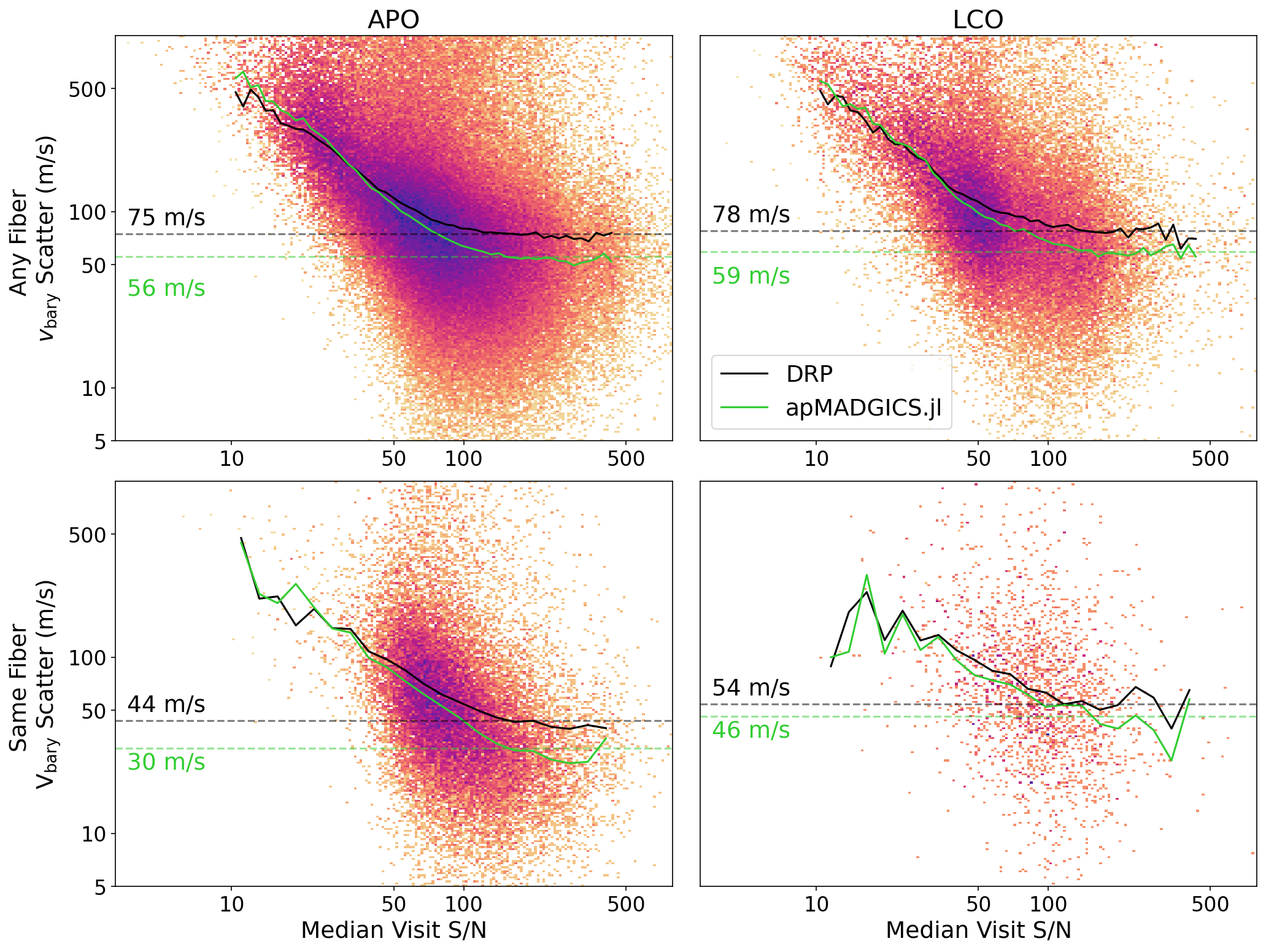}
\caption{Log-density histograms of the scatter in Barycentric velocity for repeat visits of the same star obtained by \apMADGICS{} versus the median signal-to-noise of the visits. Repeats are subdivided by telescope, APO (left) and LCO (right), and by repeat observations using any fiber (top) or restricted to observations using the same fiber (bottom). The average velocity scatter versus S/N is shown for the DRP (black solid line) and \apMADGICS{} (green solid line). The median velocity scatter at high S/N (120 - 450) is indicated by a dashed line matched in color and labeled rounded to the nearest m/s. Cuts to targets based on ASPCAP quality flags, stellar type, and $\sigma(v_{\rm{bary}}) < 1$ km/s to reject outliers were applied (see text for more).
}
\label{fig:helioScatterVersusSNR}
\end{figure*}

\pagebreak
\subsection{RV Repeat Visit Validation} \label{sec:RVRepeat}

For each visit spectrum, we infer the sky frame (of the observatory) radial velocity $v_{\rm{rel}}$ by finding the minimum $\Delta\chi^2$ as a function of pixel shift for the stellar line prior. The pixel shifts are sampled on a discrete grid of 0.1 pixel shifts and the $\Delta\chi^2$ surface is then quadratically interpolated to yield a continuous estimate of $v_{\rm{rel}}$. Because all components of the model, notably the sky line and tellurics in the star continuum, are being re-evaluated at each test velocity, we are finding the observed radial velocity while marginalizing over these other components. We then convert to the barycentric radial velocity $v_{\rm{bary}}$ using the mid-point in time for the observation and the location of each observatory.\footnote{Observatory locations were queried from \texttt{astropy} v5.3.4} There are several timing keywords in the raw data headers and while the DRP uses ``MJD-OBS'' (TAI)\footnote{The FITS header erroneously labels the ``MJD-OBS'' as being UTC, but it is actually TAI.} plus half of the ``EXPTIME'' in the ``apVisit'' files to define the midpoint of the observation, \apMADGICS{} uses the mean of all of the per exposure ``JD-MID'' in the visit. We follow \citet{Wright_2014_PASP} for the barycentric correction $z_{\rm{corr}}$, applied multiplicatively (Eq. \ref{eq:vcorr}), using the implementation in \texttt{astropy} (v5.3.4) that neglects Shapiro delay and light travel time ($\sim 1$ cm/s corrections).\footnote{\texttt{astropy} outputs $z_{\rm{corr}}*c$ and erroneously suggests applying the correction as $v_{\rm{bary}} = v_{\rm{rel}} + v_{\rm{corr}} + v_{\rm{rel}}v_{\rm{corr}}/c$. We instead divide the \texttt{astropy} output by $c$ and correctly apply the multiplicative correction as Eq. \ref{eq:vcorr} before then converting $z_{\rm{bary}}$ to $v_{\rm{bary}}$ fully relativistically.}
\begin{ceqn}
\begin{align} \label{eq:vcorr}
    z_{\rm{bary}} &= z_{\rm{rel}} + z_{\rm{corr}} + z_{\rm{rel}}z_{\rm{corr}}
\end{align}
\end{ceqn}
\indent A useful validation is to check if the inferred $v_{\rm{bary}}$ depends on the observed radial velocity $v_{\rm{rel}}$. While stellar variability introduces ``real'' scatter in the $v_{\rm{bary}}$ measurement, it should not correlate with $v_{\rm{rel}}$ because we do not expect any connection between the physics of the observed star and when it is observed. Thus, any dependence of $v_{\rm{bary}}$ on the observed radial velocity $v_{\rm{rel}}$ would indicate a bias introduced by effects in the ``sky frame.'' To probe this, we compute the mean $v_{\rm{bary}}$ for stars with repeat observations and plot the observed difference from this mean as a function of $v_{\rm{rel}}$. Here we have defined a ``star'' to be a unique combination of \texttt{APOGEE\_ID} and \texttt{TELESCOPE}.\footnote{Some previous analyses of the RV precision of APOGEE have restricted to considering repeat visits of the same plate, which artificially suppresses scatter associated with fiber positioning and longer time baselines.} This means that the same star observed in the north and south will be viewed as a different object for this test. We apply a cut to only consider stars with more than 5 good RV visit measurements from both the DRP and \apMADGICS{}. The results are shown in the left and middle panels of Figure \ref{fig:RV_DRP_MAD}, respectively.

The comparison for the DRP shows coherent deviations in the outlier population of the per visit $v_{\rm{bary}}$ compared to the mean $v_{\rm{bary}}$ as a function of $v_{\rm{rel}}$, while these deviations are absent for \apMADGICS{} (Figure \ref{fig:RV_DRP_MAD}, top row). This suggests that \apMADGICS{} effectively marginalizes over ``sky frame'' effects to produce more unbiased measurements of the radial velocity of stars. The narrower y-axis range in the bottom panel reveals these overdensities of outliers at specific $v_{\rm{rel}}$ are present even for visits with small differences relative to the mean. The deviations/overdensities are even more apparent in the difference between the DRP and \apMADGICS{} $v_{\rm{rel}}$ versus \apMADGICS{} $v_{\rm{rel}}$, which requires no cuts on repeat number of visits (see Figure \ref{fig:RV_DRP_MAD}, right panel).

For $\sim4$\% of the RVs, the DRP and \apMADGICS{} disagree significantly, by more than 20 km/s. While most are simply in the extremely low S/N limit, we investigated and visualized several sub-populations of these outliers, where the results from both pipelines are likely incorrect. One population comes from low S/N spectra that also contain a large, unmasked, recurring outlier pixel/artifact near 15650 \r{A}, which \apMADGICS{} tends to assign as RV $+200$m/s. Other common cases are carbon and oxygen-rich AGB stars in the LMC, to which the DRP tends to assign RVs 50-100 km/s larger than \apMADGICS{} does. Investigating these outliers/failure-modes to improve our detector/artifact masks and the efficacy of \apMADGICS{} priors for unusual stellar types will be examined in future work.

\section{RV Precision} \label{sec:RVPrecision}

For stars with little intrinsic RV variability, measuring the spread of these repeat measurements of $v_{\rm{bary}}$ for the same star provides a measure of the floor on our radial velocity precision due to systematics. We restrict the stellar type of targets, require they be well-modeled by ASPCAP, and have $\sigma(v_{\rm{bary}}) < 1$ km/s (see below and Equation \ref{eq:scatter1}) to reject outliers (see Section \ref{sec:Data}). We also restrict to stars with at least 3 good RV visits so we can calculate the spread $\sigma(v_{\rm{bary}})$ with the standard deviation of $v_{\rm{bary}}$ using jackknife resampling to help offset biases introduced by the small sample sizes ($\sim50\%$ of all stars with repeat visits have $\leq5$ visits). For an idea of magnitude, using a simple standard deviation without jackknifing leads to 17\% and 11\% underestimation of the scatter for sample size $N=3$ and $N=4$, respectively.

Jackknife resampling is a statistical technique to estimate and correct the bias of given statistic $f(x_1,...,x_N)$ on a fixed sample of size $N$ by computing that statistic over all subsets of the sample that leave out one observation. By comparing the mean of that statistic over all subsets $\widetilde{\theta}_{\rm{jack}}$ (Equation \ref{eq:jack_average}) with the statistic on the full dataset ($\hat{\theta}$), one can estimate and correct the estimate of that statistic $\hat{\theta}_{\rm{jack}}$ \citep[Equation \ref{eq:jack_bias_cor}, ][]{berger2006adjusted}. Similarly, one can estimate the variance of the jackknifed estimate $\hat{\sigma}^2(\hat{\theta}_{\rm{jack}})$ for a statistic given the variance of that statistic on the leave-one-out subsets $\tilde{\sigma}_{\rm{jack}}^2(\theta)$ \citep[Equation \ref{eq:jack_var},][]{berger2005jackknife}.

\begin{ceqn}
\begin{align} \label{eq:jack_average}
    \widetilde{\theta}_{\rm{jack}} &= \frac{1}{N} \sum_{i=1}^{N} f(x_1,...,x_{i-1},x_{i+1},...,x_N) \\
    \label{eq:jack_bias_cor}
    \hat{\theta}_{\rm{jack}} &= N\hat{\theta}-(N-1)\widetilde{\theta}_{\rm{jack}} \\
    \label{eq:jack_var}
    \hat{\sigma}^2(\hat{\theta}_{\rm{jack}}) &= \frac{\left(N-1\right)^2}{N}\tilde{\sigma}_{\rm{jack}}^2(\theta)
\end{align}
\end{ceqn}

In our first analysis (Figure \ref{fig:helioScatterVersusSNR}), we use Equation \ref{eq:scatter1} as the statistic that we jackknife, which most closely corresponds to the ``usual'' notion of precision, the scatter of repeat measurements around the sample mean. However, in the presence of fiber-fiber offsets, this sample mean is biased towards the average fiber used to measure a given star. It is important to understand this scatter to, for example, understand the systematics introduced by measuring a single star using multiple fibers when doing stellar activity/flare detection. Scatter due to fiber-fiber offsets is also important when trying to describe the precision on the relative velocity between two stars, up to gravitational/convective shifts, that were measured with different fibers. To describe this second notion of precision, we use Equation \ref{eq:scatter2} as the statistic which we jackknife, where we subtract the inverse-variance weighted mean of the fiber-offset corrected RVs. We discuss these corrections and resulting precision estimates in Section \ref{sec:RV_rel_fib_cal}.
\begin{ceqn}
\begin{align} \label{eq:scatter1}
    \sigma(v_{\rm{bary}}) &= \sqrt{\frac{1}{N-1} \sum_{i=1}^N \left( x_{i} - \frac{1}{N}\sum_{j=1}^N x_{j} \right)^2} \\
    \label{eq:scatter2}
    \sigma(v_{\rm{bary}}) &= \sqrt{\frac{1}{N-1} \sum_{i=1}^N \left( y_{i} - \frac{1}{S}\sum_{j=1}^N w_{j} y_{j} \right)^2}
\end{align}
\end{ceqn}
Here N is the total number of observations of a given star and the $x_i$ are individual epoch, solar barycenter-corrected RV measurements. In Equation \ref{eq:scatter2}, $w_i$ are inverse-variance weights using the RV errors after calibration described in Section \ref{sec:RVcal}, S is the sum of the weights, and $y_i$ are individual epoch, solar barycenter-corrected, fiber-offset corrected RV measurements.

In the top row of Figure \ref{fig:helioScatterVersusSNR}, we show this spread $\sigma(v_{\rm{bary}})$ versus the median S/N of the repeat visits as a 2D histogram with logarithmic color scale for density, separated by telescope (APO left and LCO right). The median $\sigma(v_{\rm{bary}})$ as a function of S/N is also plotted for the DRP and \apMADGICS{}, black and green lines respectively. The median $\sigma(v_{\rm{bary}})$ for all high S/N measurements (120 $<$ S/N $<$ 450) is plotted as a horizontal dashed line, and represents an estimate of the systematics floor of the radial velocity measurement. Thus, \apMADGICS{} achieves a precision of 56 m/s and 59 m/s at APO and LCO, respectively, without correcting for fiber-fiber RV offsets. 

We expect slightly lower velocity precision at LCO as the result of the wider LSFs compared to APO, reducing the effective resolution of the instrument (see Appendix \ref{sec:LSFCentroid}). For example, using the method described in Section \ref{sec:RVType}, the expected difference in photon-noise limited RV precision for the average fiber at APO versus LCO is 1.3 m/s at S/N 50 for a star with ($T_{\rm{eff}}$, $\log g$, [X/H]) = (4523 K, 2.43, -0.21), an example which is described in Appendix \ref{sec:QfactorChips}. The difference in RV precision between APO and LCO could also be caused by a difference in the distribution of stellar types targeted by each instrument or the relative stability of their wavelength solution, which bears further investigation. Overall, the comparison of RVs from \apMADGICS{} to previous measurements from the DRP shows that \apMADGICS{} achieves a $\sim19$ m/s improvement, without including fiber-fiber constant offsets.

\begin{figure*}[t]
\centering
\includegraphics[width=\linewidth]{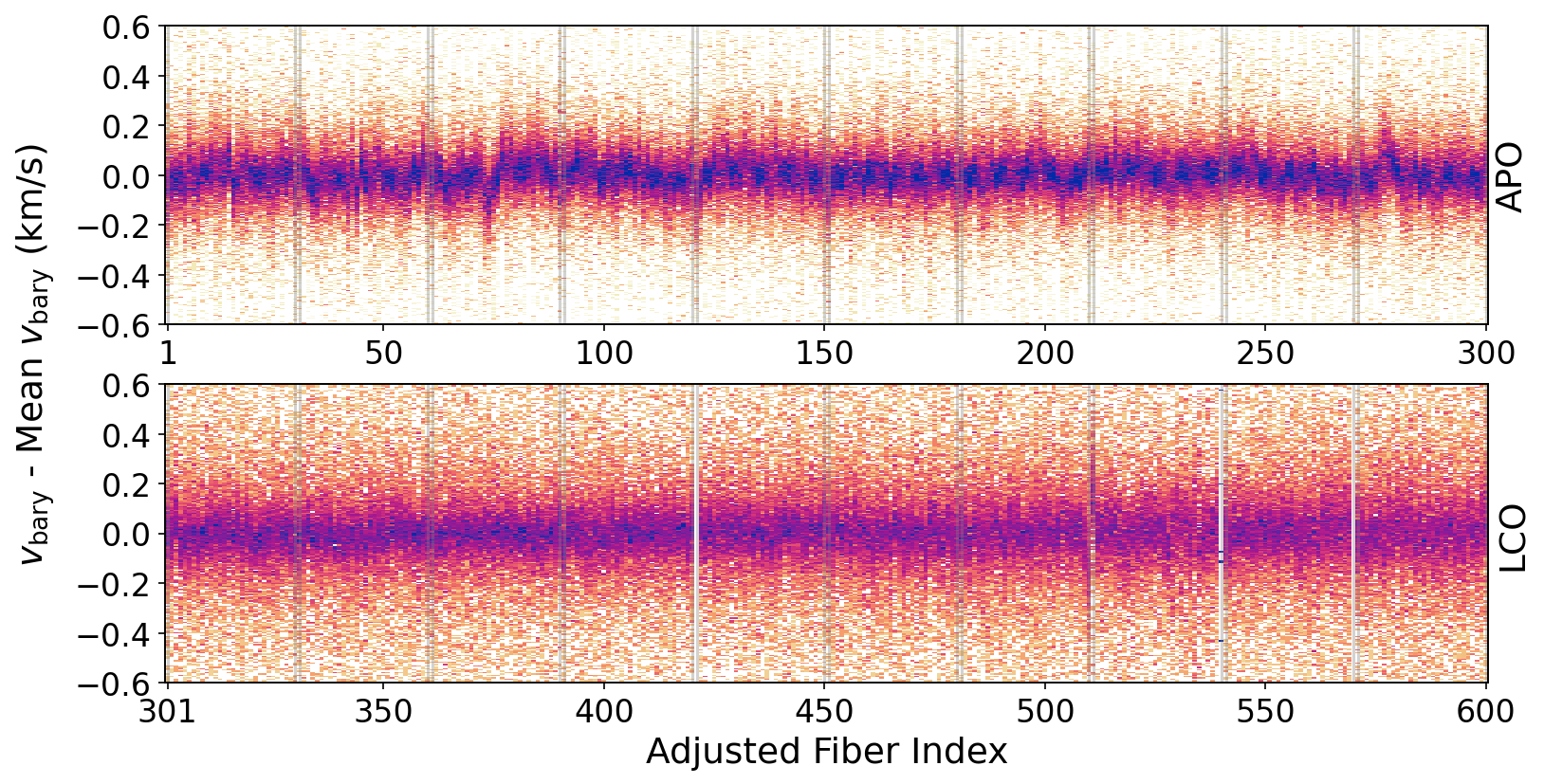}
\caption{For stars with at least 5 good RV visits, the deviation of the barycentric velocity obtained for a single visit from the average barycentric velocity for the star is shown as a 2D histogram versus the fiber used for that visit (top panel are APO fibers, bottom panel are LCO fibers). The 2D histogram is column normalized and the color is logarithmic. Guidelines (grey) indicate the edges of the physical fiber blocks.
}
\label{fig:RV_per_fiber}
\end{figure*}

To better understand the source of the $\sigma(v_{\rm{bary}})$, we further restrict to stars where all repeat measurements were obtained with the same fiber (Figure \ref{fig:helioScatterVersusSNR}, bottom row). This allows us to remove systematics caused by the variation in LSF between fibers. Restricted to the same fiber, \apMADGICS{} achieves 30 m/s and 46 m/s precision at APO and LCO, respectively. However, there are few repeat measurements at LCO with the same fiber which makes estimating the precision for LCO difficult in this case. Even when restricted to observations taken with the same fiber, we still observe an improvement over the DRP radial velocity of $\sim14$ m/s. Because uncertainties add in quadrature, and the RVs are so precise, the $14$ m/s improvement at APO corresponds to reducing the systematics by half ($44^2 - 30^2 = 32^2$). For context, this 30 m/s RV precision means \apMADGICS{} is measuring shifts in spectra comparable to 6/1000$^{\rm{th}}$ of the physical pixel size.

To further confirm that fiber-to-fiber variability of the LSF is causing a significant increase in the scatter of our radial velocities, we consider all stars well-modeled by ASPCAP and with $\geq5$ visits. We plot the deviation between the $v_{\rm{bary}}$ measured for a single visit to the mean $v_{\rm{bary}}$ for a given star, as a function of adjusted fiber index (Figure \ref{fig:RV_per_fiber}). Especially for APO, clear fiber dependent offsets in $v_{\rm{bary}}$ are observed, showing spatial coherence (since adjacent fiber indices are spatially adjacent on the detector). This motivates deriving at least a fiber-dependent constant RV offset to help mitigate this fiber-fiber variation. 

\begin{figure*}[t]
\centering
\includegraphics[width=\linewidth]{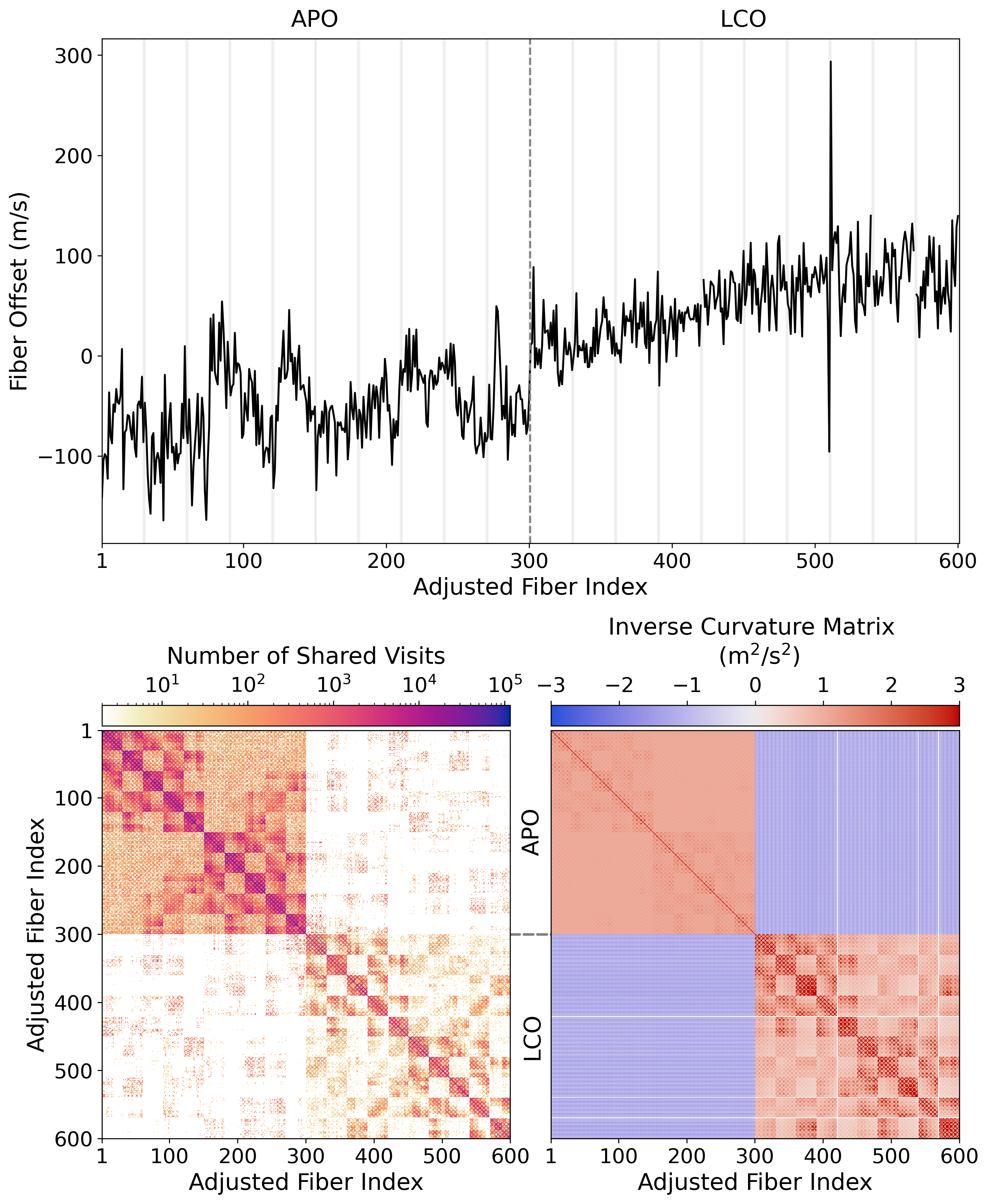}
\caption{Top Panel: Constant RV offset as a function of fiber number. Dashed line divides fiber axis between the two telescopes, light gray lines show the boundaries between physical v-groove blocks of 30 fibers. Bottom Left: Adjacency matrix showing number of visits to stars in common between two different fibers. Patchwork pattern is partially due to fiber blocks, which is discussed further in Section \ref{sec:RV_rel_fib_cal}. Bottom Right: Inverse curvature matrix representing uncertainty in the estimation of the per fiber RV offsets.
}
\label{fig:fiberOffsets}
\end{figure*}

\section{RV Catalog Calibration} \label{sec:RV_fib_cal}

\subsection{Relative Fiber Calibration} \label{sec:RV_rel_fib_cal}
In order to determine the per-fiber constant RV offset, we leverage stars with multiple visits, where at least two of those visits use different fibers. Then, we set up and solve a linear system of equations for the offsets (assumed to be constant in time), minimizing the variance of repeat measurements. Such an approach is often called ``ubercalibration''  \citep{Padmanabhan:2008:ApJ:,Schlafly:2012:ApJ:,Magnier:2013:ApJS:,Magnier:2020:ApJS:,Schlafly:2018:ApJS:,Saydjari_2023_ApJS}. We solve this generally overdetermined linear problem ($A\vec{x}=\vec{b}$) by minimizing the residuals ($r = A\vec{x}-\vec{b}$) with respect to measurement uncertainty (expressed as the inverse covariance matrix $C^{-1}$) for the parameters $\vec{x}$, $\chi^2 = (A\vec{x}-\vec{b})^TC^{-1}(A\vec{x}-\vec{b}) = r^TC^{-1}r$. We make the common approximation that each measurement/observation is independent, and so take $C^{-1}$ to be diagonal. However, this independence assumption is violated by instrumental and environmental effects that introduce spatial and temporal correlations.

First, we will describe a model for solving the ``full'' problem, and then describe the model we use to solve for only the fiber-fiber constant RV offsets. For the ``full'' problem, one solves for both the fiber offsets and the unknown true RV of each star, which we are implicitly assuming is time invariant after solar barycenter corrections. This is a $\chi^2$ minimization over $RV_{i}$ and $f_{j}$ for the residual given in Equation \ref{eq:sparseCal}.
\begin{ceqn}
\begin{align} \label{eq:sparseCal}
    r =  \left( RV_{i} + f_{j} \right) - \widetilde{RV}^k_{ij}
\end{align}
\end{ceqn}
Here $\widetilde{RV}^k_{ij}$ is a measured RV for star $i$ with fiber $j$ and visit index $k$, $RV_{i}$ is the ``true'' radial velocity of star $i$ (assumed constant during the survey), and $f_{j}$ is an RV offset for fiber $j$. The fiber index $j$ implicitly depends on the visit index $k$. The ``design matrix,'' A in the linear system $A\vec{x}-\vec{b}$, is $n_{\rm{obs}} \times (n_{\rm{stars}} + n_{\rm{fibers}})$. Each row corresponds to a visit/observation $\widetilde{RV}^k_{ij}$ and has a 1 in both the $i$-th and $n_{\rm{stars}}+j$-th column. The parameters vector $\vec{x}$ is $n_{\rm{stars}} + n_{\rm{fibers}}$ long, containing both the 600 fiber offsets of interest and the RV estimates for unique stars, $\sim200,000$ in this case. Even with our large problem size of $\sim1.2$ million observations (the entries in $\vec{b}$), this can be solved for $\vec{x}$ efficiently using sparse orthogonal-upper triangular (QR) or lower-upper triangular (LU) factorization, for example using \texttt{LinearSolve.jl}. Unfortunately, obtaining the inverse curvature matrix, which provides access to estimates of the parameter uncertainties, is not similarly tractable in this formulation. 

We instead solve a more limited model for just the the $f_j$ by fixing the $RV_{i}$ subspace to the optimal point. The maximum likelihood estimate for $RV_{i}$ is equal to the inverse-variance weighted mean of all fiber-corrected observations, $\widetilde{RV}_{ij}-f_{j}$, for a given star $i$. 
\begin{ceqn}
\begin{align} \label{eq:subRVi}
    \widehat{RV}_{i} &= \frac{1}{S_i} \sum_m \left(\widetilde{RV}^m_{ij} - f_j\right)w^m
\end{align}
\end{ceqn}
Here $w$ is the inverse variance weight of a visit given by the uncertainty in the RV measurement, $S_i$ is the sum over the weights for all visits for star $i$, and the sum over $m$ is over all visits to star $i$. By substituting Equation \ref{eq:subRVi} into \ref{eq:sparseCal}, we eliminate $n_{\rm{stars}}$ parameters from the model at the cost of making the design matrix slightly more difficult to construct. Now the $\chi^2$ minimization is only over the $f_j$.

\begin{ceqn}
\begin{align} \label{eq:finalModel}
   r = \frac{1}{S_i} \sum_m \left(\widetilde{RV}^m_{ip} - f_p \right)w^m + f_{j} - \widetilde{RV}^k_{ij}
\end{align}
\end{ceqn}
\begin{ceqn}
\begin{align} \label{eq:designmat}
\begin{split}
     r =  \left(f_j - \frac{1}{S_i} \sum_{m} f_{p}w^m\right) - \\ \left(\widetilde{RV}^k_{ij} - \frac{1}{S_i} \sum_m \widetilde{RV}^m_{ip}w^m\right)
\end{split}
\end{align}
\end{ceqn}
The left term in Equation \ref{eq:designmat} specifies the design matrix, which is now $n_{\rm{obs}} \times n_{\rm{fibers}}$. For each observation (row), a one is placed in the $j$-th column and then the total fractional weight per fiber of visits for that star are subtracted from the relevant columns. The right term in Equation \ref{eq:designmat} indicates that, under this model, $\vec{b}$ is the observed visit radial velocity minus the mean over all other visits of that star. This formulation makes solving for both the parameters $\vec{x}$ and the inverse curvature matrix tractable with direct matrix inversion, such as singular-value decomposition (SVD), because the number of parameters is only $n_{\rm{fibers}}$, $\sim 600$ in our case.

\begin{ceqn}
\begin{align} \label{eq:lsqs}
     \vec{x} &= (A^TC^{-1}A)^{-1}A^TC^{-1}\vec{b} \\
     \left(\frac{\partial^2\chi^2}{\partial x_i \partial x_j}\right)^{-1} &= (A^TC^{-1}A)^{-1} \label{eq:lsqs_err}
\end{align}
\end{ceqn}
Here $C^{-1}$ is generally approximated as a diagonal matrix with the inverse variances associated with the uncertainties on each observation on the diagonal. Unlike Equation \ref{eq:sparseCal}, Equation \ref{eq:finalModel} depends on more than one observation per row. Thus, even if the observations were truly independent, taking $C^{-1}$ to be diagonal ignores the correlations introduced by the sum. This means that the uncertainties from Equation \ref{eq:lsqs_err} will be strictly underestimated under this approximation for $C^{-1}$.

Because the systematic corrections to the RV uncertainty will depend slightly on the derived fiber-fiber offset corrections (see Section \ref{sec:RVcal}), we break this potential circularity by first solving the unweighted problem ($C^{-1} = I$), then estimating the systematics-corrected RV uncertainties, and re-solving for the fiber offsets using those uncertainties in $C^{-1}$.

\begin{figure*}[t]
\centering
\includegraphics[width=\linewidth]{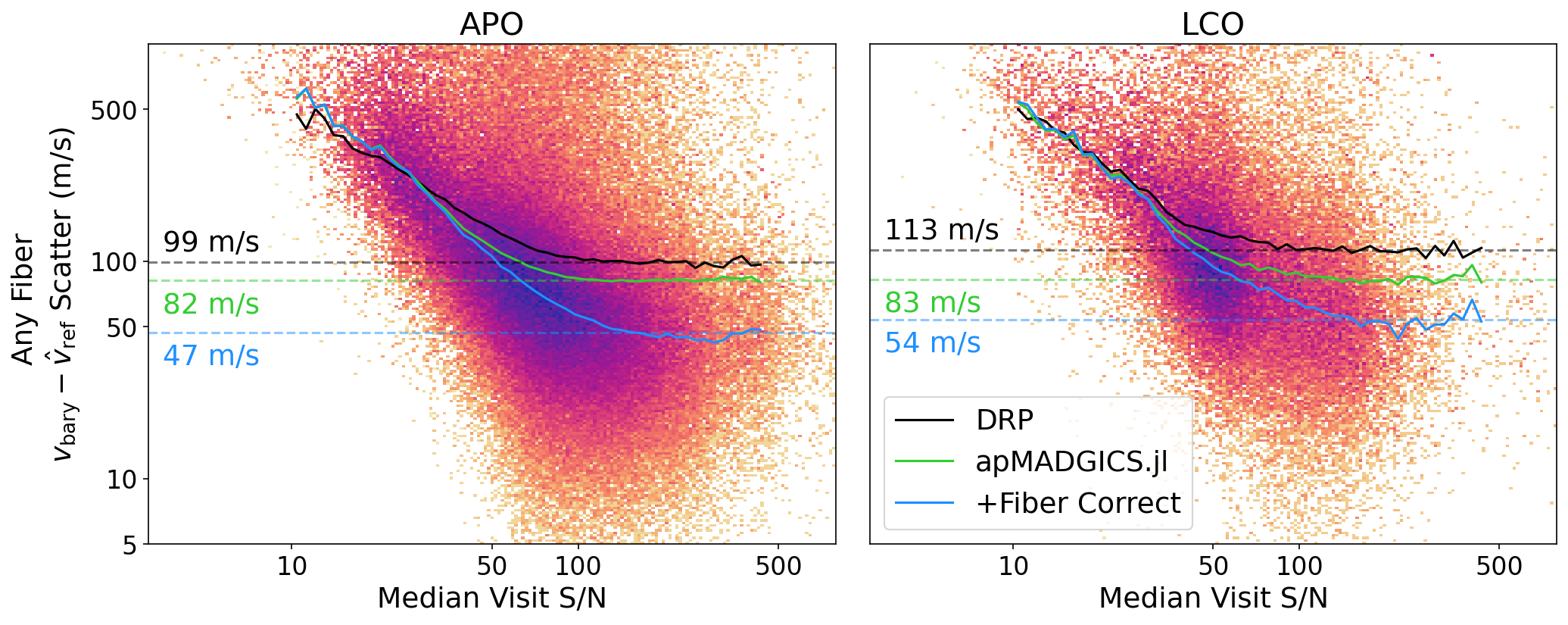}
\caption{Same as Figure \ref{fig:helioScatterVersusSNR}, but with the scatter measured relative to the maximum likelihood estimate for the ``true'' RV for each star, $\hat{v}_{\rm{ref}}$. The average velocity scatter versus S/N is shown for the DRP (black solid line), \apMADGICS{} (green solid line), and \apMADGICS{} after correcting for constant RV fiber-fiber offsets (blue solid line).
}
\label{fig:RV_prec_postCorr}
\end{figure*}

Because the absolute zeropoint is entirely unconstrained by this analysis, there is always at least one eigenvalue in the SVD of $(A^TC^{-1}A)$ that is numerically zero. Thus, in practice, we compute $(A^TC^{-1}A)^{-1}$ as $VS^{-1}U^{T}$ where eigenvalues in $S$ below a threshold are set to zero rather than inverted. In our case, the unconstrained eigenvalue is 6-10 orders of magnitude smaller than the next largest eigenvalue, so the cutoff is clear.

Our application of this method to the \apMADGICS{} RV catalog is summarized in Figure \ref{fig:fiberOffsets}. We limit observations to visits of stars that were observed with more than one fiber, that are well-modeled by ASPCAP, with stellar types $\log(g) > 1.3$ and T$_{\rm{eff}} < 6000$, with good visit RVs from \apMADGICS{}, and reject RV outliers (see Section \ref{sec:Data}). 

After these cuts, two fibers, 421 and 570 (LCO FIBERID 180 and 31), have no remaining visits and fiber 540 (LCO FIBERID 61) only has 2 ``good'' RV visits. Thus, we view all 3 fibers as unable to be calibrated, exclude them from this and further analysis, and set RVs measured with these fibers to NaN. Even without these cuts, all visit spectra with fibers 421 and 570 are either NaN-ed out upstream by the DRP or by \apMADGICS{} because they are below the throughput threshold beyond which fluxing from the DRP was determined to be unstable. For fiber 540, only 35 visits (2.6\%) are above this throughput threshold. For context, 12\% (fiber 421), 75\% (fiber 570), and 5\% (fiber 540) of those RVs from the DRP are NaNs. However, even after excluding these fibers, only 1\% of all visits have NaN RVs from \apMADGICS{} compared to 3\% from the DRP, so \apMADGICS{} strictly improves the RV success rate despite excluding these fibers. Some low-throughput mitigation was performed between SDSS-IV and SDSS-V, though only at APO \citep{Wilson_2022_SPIE_reterm}.

We present a compressed summary of the fiber-fiber connectivity through shared observations of the same star in the bottom left of Figure \ref{fig:fiberOffsets} as an adjacency matrix. Each entry represents the number of visits to a star observed with both fiber $m$ and $n$. There are three main layers of structure in this adjacency matrix. First, there is little overlap between APO and LCO in targets, in large part because they are in different hemispheres. Within each telescope, there are 10 clear blocks of 30 fibers that correspond to the physical fiber bundling, which placed constraints on the interchange of fibers during plugging. Within each block of 30 fibers, there is also a period on the scale of 6 fibers associated with cross-talk and persistence mitigation that caused each fiber to be assigned as for either ``faint'', ``medium'', or ``bright'' targets. These fibers were then ordered on the detector as f-m-b-b-m-f, which is discussed in more detail in \citet{Majewski_2017_AJ}. There also appears to be a slight exclusion between observations of adjacent fiber bundles and a strong exclusion between the first five and last 5 fiber bundles, especially at APO. These inter-fiber bundle patterns likely reflect the mapping of fiber bundle order on the slit head to the location of those bundles on the focal plane.

The constant RV offsets per fiber are shown in the top panel of Figure \ref{fig:fiberOffsets}. The mean offset between APO and LCO is 104.2 m/s. The largest amplitude corrections are for fibers 510 and 511 at LCO (LCO FIBERID 91 and 90). The spatially coherent structures in the offsets tend to be largest in amplitude near the edges of the physical v-groove blocks of 30 fibers (light gray lines) or have periodicity roughly matched to that size. There is significantly less structure in the offsets at LCO, which may reflect the improved alignment of the v-groove blocks on the pseudo-slit at LCO. However, we note that if the wavelength solution and LSFs were sufficiently flexible and well-constrained, these features should not appear. Thus, we can take the amplitude and structure of these corrections as an indicator of LSF and wavelength solution performance in further APOGEE software development in SDSS-V. 

As mentioned in Section \ref{sec:RVPrecision}, the analysis in Figure \ref{fig:helioScatterVersusSNR} underestimates systematics due to fiber-fiber RV offsets because there we computed standard deviation after first subtracting the sample mean. However, the sample mean tends to be biased toward the fiber used most frequently to measure a given star. If we want to understand the precision of relative RVs between stars measured with different fibers in the fiber-merged RV catalog, we should measure the scatter around the maximum likelihood estimate for the ``true'' radial velocity of a star, which is the inverse-variance weighted, constant per fiber RV offset corrected, radial velocity, which is what we show in Figure \ref{fig:RV_prec_postCorr}. To improve the comparison to the DRP results, we have subtracted a single constant from all stars in the DRP catalog to account for any offsets in the absolute RV calibration (see Section \ref{sec:RV_abs_fib_cal}). We chose this constant to be the median difference between the DRP RVs and the \apMADGICS{} maximum likelihood estimate for the ``true'' radial velocity of a star.\footnote{Because jackknife estimates of the standard deviation with mean zero can still be biased slightly low ($\sim5\%$ at $n=3$), we use mean zero centering for the catalogs without fiber-fiber offset correction and let the mean float for the fiber-corrected \apMADGICS{} catalog to provide the most conservative estimate of RV precision improvement.} This illustrates that, with respect to this notion of precision, the RV precision for both the DRP and \apMADGICS{} was worse than the simple estimate given in Figure \ref{fig:helioScatterVersusSNR}, only 99 m/s and 82 m/s at APO, respectively.

After applying the fiber-fiber RV offsets, we improve the RV precision of the catalog pooled over fibers to 47 m/s at APO and 54 m/s at LCO (blue lines in Figure \ref{fig:RV_prec_postCorr}). In total, this allows us to achieve a 41 m/s (59 m/s) improvement in RVs at APO (LCO) compared to the DRP for the fiber-merged catalog. Because of how we defined the centering, these fiber-corrected RV precisions can also be directly compared to the top panel of Figure \ref{fig:helioScatterVersusSNR}, if the usual definition of RV precision is desired, and still show improvements from 56 m/s to 47 m/s (59 m/s to 54 m/s) at APO (LCO). However, the fact that our fiber-offset corrected RV catalog is still not achieving the same RV precision as repeat observations using a single fiber indicates that further improvements to the LSF and wavelength solutions would be productive. We discuss one possible cause for fiber-fiber variations in RVs associated with the wavelength calibration procedure in Appendix \ref{sec:LSFCentroid}.

We show the inverse curvature matrix, which provides an estimate of the uncertainty on the recovered parameters, in the bottom right of Figure \ref{fig:fiberOffsets}. This would be exactly the formal covariance matrix of the parameters if we had solved the ``full'' problem (Equation \ref{eq:sparseCal}) or accounted for the correlations between rows (visits) introduced by Equation \ref{eq:finalModel} in $C^{-1}$. The structure in this matrix illustrates the correlated uncertainties introduced by the structure in the adjacency matrix. Specifically, there is a strong anti-correlation between RVs observed at APO and LCO, which highlights the lack of a tight constraint on the offset between the two telescopes given the relatively few stars observed by both telescopes. Quantitatively, the square-root of the diagonal of the inverse curvature matrix can be interpreted as approximate uncertainties in the parameter estimation, ignoring off-diagonal correlations.

For APO this is 2.7, 2.0, and 1.9 m/s for the faint, medium and bright fibers, is 4.6, 4.0, and 3.9 m/s for LCO. The mean APO-LCO offset uncertainty estimate is 2.1 m/s and is obtained from the off-diagonal elements by the inner product of the inverse curvature matrix with a vector that is $1/n_{\rm{APO}}$ for the APO fiber indices and $-1/n_{\rm{LCO}}$ for the LCO fiber indices. These uncertainties represent a lower bound on the uncertainties of the calibration based solely on the structure of the shared visits and measurement uncertainties. Systematics like time dependent drift in the RVs and scatter from RV variability will degrade our constraints on these fiber-fiber offsets. However, these uncertainties suggest that our constraints on the fiber-fiber RV offsets are small relative to the amplitude of the offsets we have found ($\sim40$ m/s) and are subdominant in our overall RV precision budget for precisions of $30-47$m/s for now. However, the uncertainties are at the scale where engineering observations to improve this calibration in the future may be valuable.

Because the Milky Way Mapper survey in SDSS-V has fewer repeat visits and robotic fiber positioners remove fiber-plugging ambiguities that improved connectivity in the adjacency matrix within the same fiber bundle, the calibration of fiber-fiber RV offsets will likely be even more challenging for SDSS-V. Thus, the ability to extract predictions using only projected visits and per-visit uncertainties before observations are complete provides valuable feedback for guiding the survey design.

\begin{figure}[t]
\centering
\includegraphics[width=\linewidth]{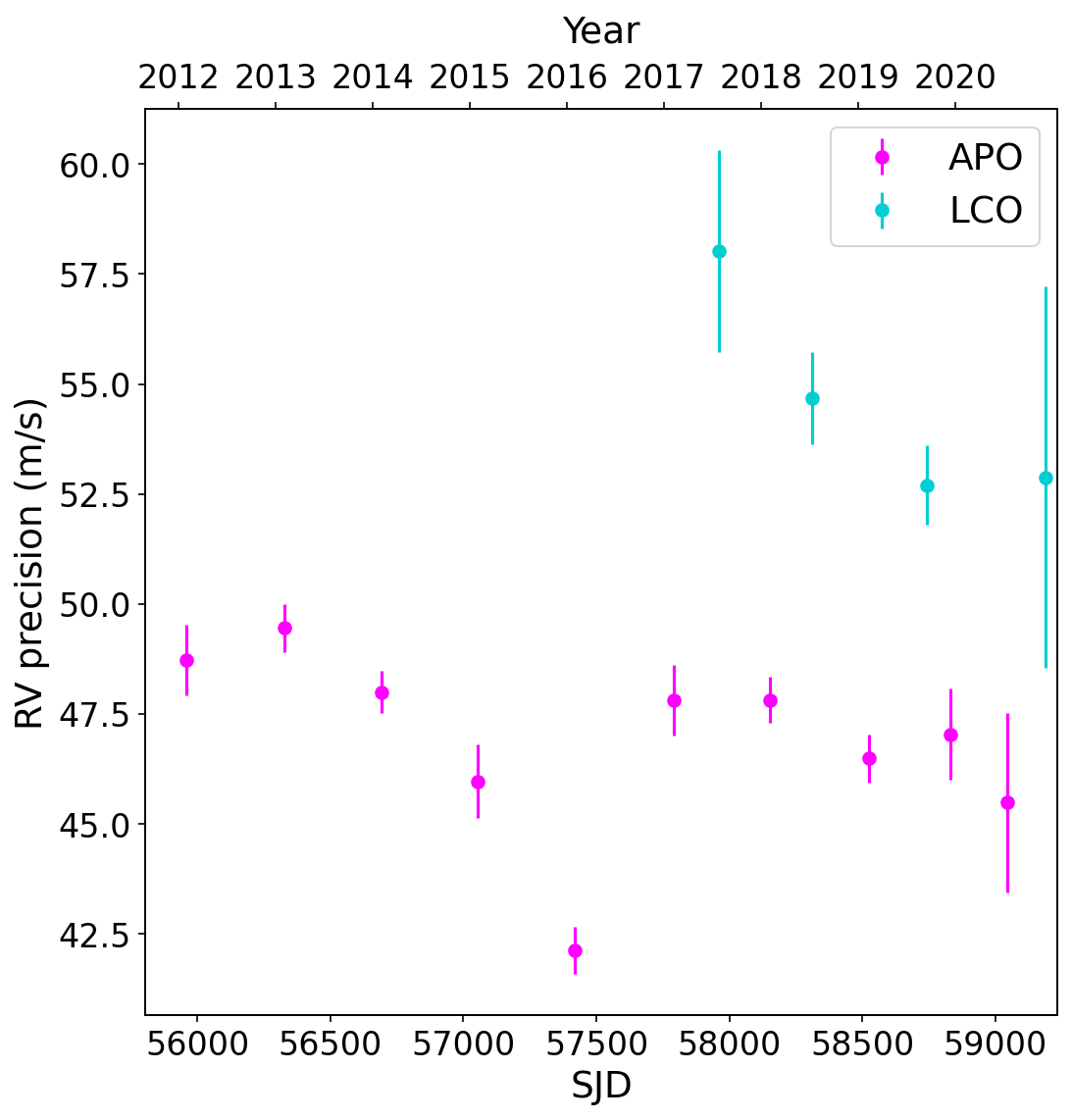}
\caption{Average RV precision per observing run.
}
\label{fig:timeDepPrec}
\end{figure}

With this large, now fiber-offset-corrected RV catalog, we can also investigate the time-dependence of the RV precision of APOGEE. We simply measure the systematic floor in the same way as Figure \ref{fig:RV_prec_postCorr}, but restricting to only those visits within each ``run'' of APOGEE (Figure \ref{fig:timeDepPrec}). For APO, these are defined as the time between shutdowns. For LCO, we chose subdivisions related to telescope and hardware improvements, the COVID shutdown and thermal cycling, and a gap in the telescope schedule to more uniformly partition the data. Error bars are estimated by jackknifing over visits using a mean estimator rather than a median and are likely to underestimate systematic uncertainties. At APO there is a clear improvement towards an optimum RV precision in Run 5, after which performance degraded to the average value and then slowly improved over the last 5 runs. At LCO, there was a significant improvement during the first few runs, which could be associated with significant improvements that were performed to the guiding hardware. 

\begin{figure*}[t]
\centering
\includegraphics[width=\linewidth]{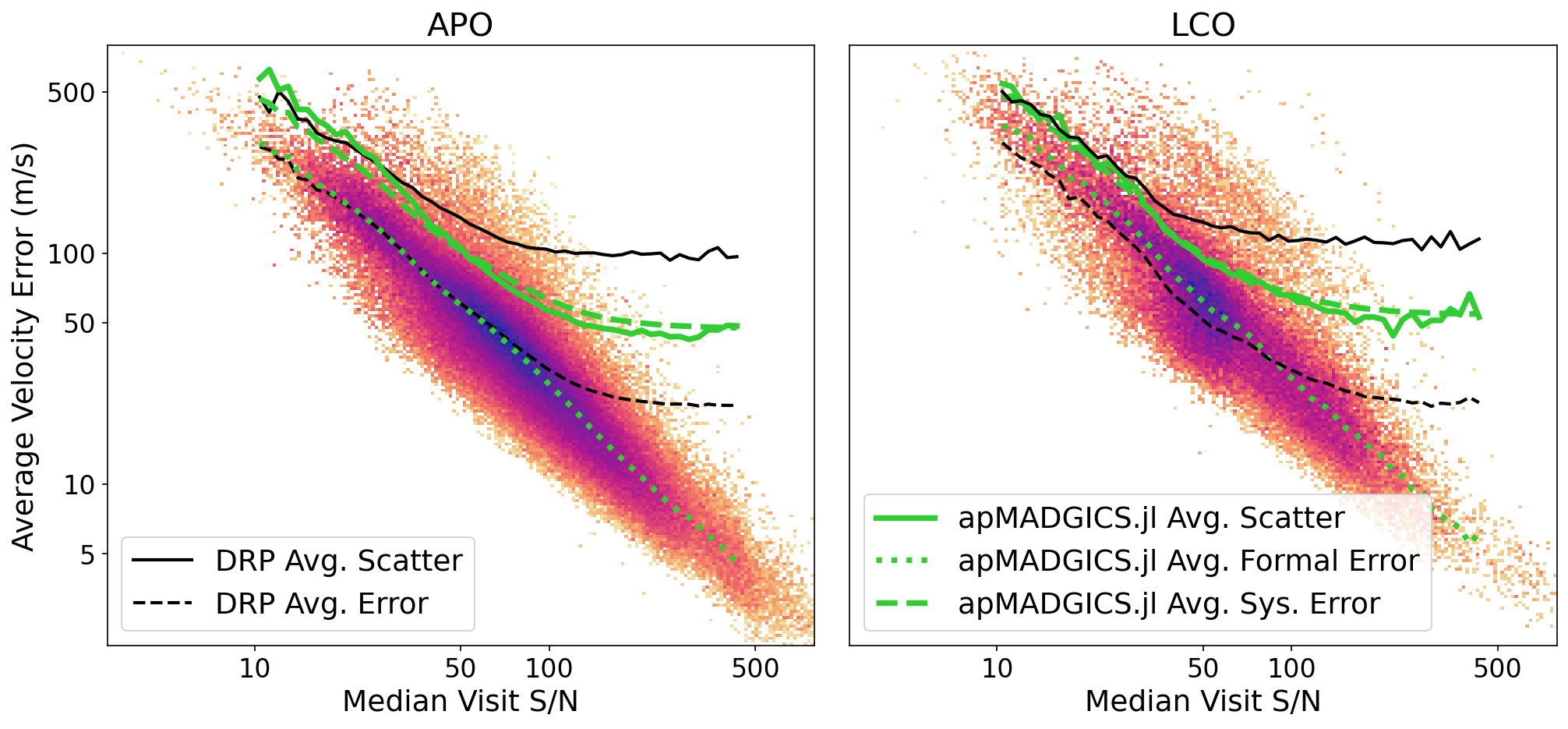}
\caption{Log-density histograms of formal radial velocity errors from \apMADGICS{} versus the median signal-to-noise of the visits of a given star for observations at APO (left) and LCO (right). The average formal error is shown as a green dotted line and the average observed scatter for repeat measurements as a green solid line. The average ``calibrated'' RV errors for \apMADGICS{} are shown as a dashed green line, and agree well with the observed scatter from repeats. The average reported error from the DRP is shown as a dashed black line and average observed scatter for repeat measurements from the DRP is shown as a solid black line.
}
\label{fig:RVsystematicsCal}
\end{figure*}

\subsection{Absolute RV Calibration} \label{sec:RV_abs_fib_cal}

Because we placed no prior constraints on the RV of stars during the relative calibration in Section \ref{sec:RV_rel_fib_cal}, that analysis cannot constrain the overall RV zeropoint. We instead tie our RV zeropoint to that given by the \korg{} models \cite{korg1,korg2} used in the \apMADGICS{} stellar prior by enforcing that the fiber-fiber constant RV correction be mean zero.

To check our RV zeropoint for external consistency, we appeal to a set of common FGK stable RV standard stars \citep{Chubak_2012_arXiv}, which are stable at the $\sim30$ m/s level. We map this catalog of 2046 stars to Gaia DR3 IDs using SIMBAD \citep{Wenger_2000_A&AS} and find 73 stars with good RV visits observed in APOGEE DR17. For each star, we take the difference between the mean of all good RV visits with \apMADGICS{} and the RV from \cite{Chubak_2012_arXiv}. 

The median difference is 558 m/s, with a scatter of 189 m/s.\footnote{Here we use the inter-quartile (IQR) range divided by the width of the IQR for a standard normal distribution as a robust measure of scatter.} For comparison, using 40 stars for APOGEE DR12, \cite{Nidever_2015_AJ} found an offset of 355 m/s with a scatter of 192 m/s. The offset between \apMADGICS{} and the DRP for DR17 is 100 m/s with 79 m/s scatter for those same stable stars, and 125 m/s with 148 m/s scatter over all stars in the catalog. Considering that this work used different stellar model spectra (from \korg{}) and derived fiber-dependent RV offsets, this agreement is well within expectations and not more than 3$\times$ the scatter. 

\section{RV Error Calibration} \label{sec:RVcal}

We also use measurements of $\sigma(v_{\rm{bary}})$ to empirically correct the \apMADGICS{} RV error bars. The average RV error estimates for the repeat visits output initially by \apMADGICS{} are shown as a 2D histogram (logarithmic color scale) versus the median S/N of the visits in Figure \ref{fig:RVsystematicsCal}. The median value as a function of S/N is shown as a green dotted line. To correct the high S/N RV uncertainties, we add the RV precision from Figure \ref{fig:RV_prec_postCorr} in quadrature with the RV errors from \apMADGICS{}. As a function of S/N, we compute the ratio of $\sigma(v_{\rm{bary}})$ on repeats to the average RV errors (ratio of the green solid and dotted lines). This ratio is approximately stable for S/N $< 120$ and we use it to derive a multiplicative correction to the initial \apMADGICS{} error bars ($1.54\times$ at APO, $1.33\times$ at LCO). This kind of multiplicative, approximately S/N independent correction can be thought of as a reduction in the number of ``effective pixels'' in the spectrum (65\% and 75\% at APO and LCO, respectively). These fractions are consistent with the fraction of pixels in the spectrum at each observatory without significant contributions from sky lines, as will be discussed in Section \ref{sec:RVType}.

By incorporating these corrections, we obtain the final RV errors reported by \apMADGICS{} (\texttt{RV\_err\_sys}), green-dashed line, which are a close match to the observed $\sigma(v_{\rm{bary}})$ for all S/N at both APO and LCO. For comparison, the $\sigma(v_{\rm{bary}})$ and average $v_{\rm{bary}}$ errors reported by the DRP are shown in black (dashed and solid, respectively).

\section{RV Stellar Type Dependence} \label{sec:RVType}

\begin{figure*}[t]
\centering
\includegraphics[width=\linewidth]{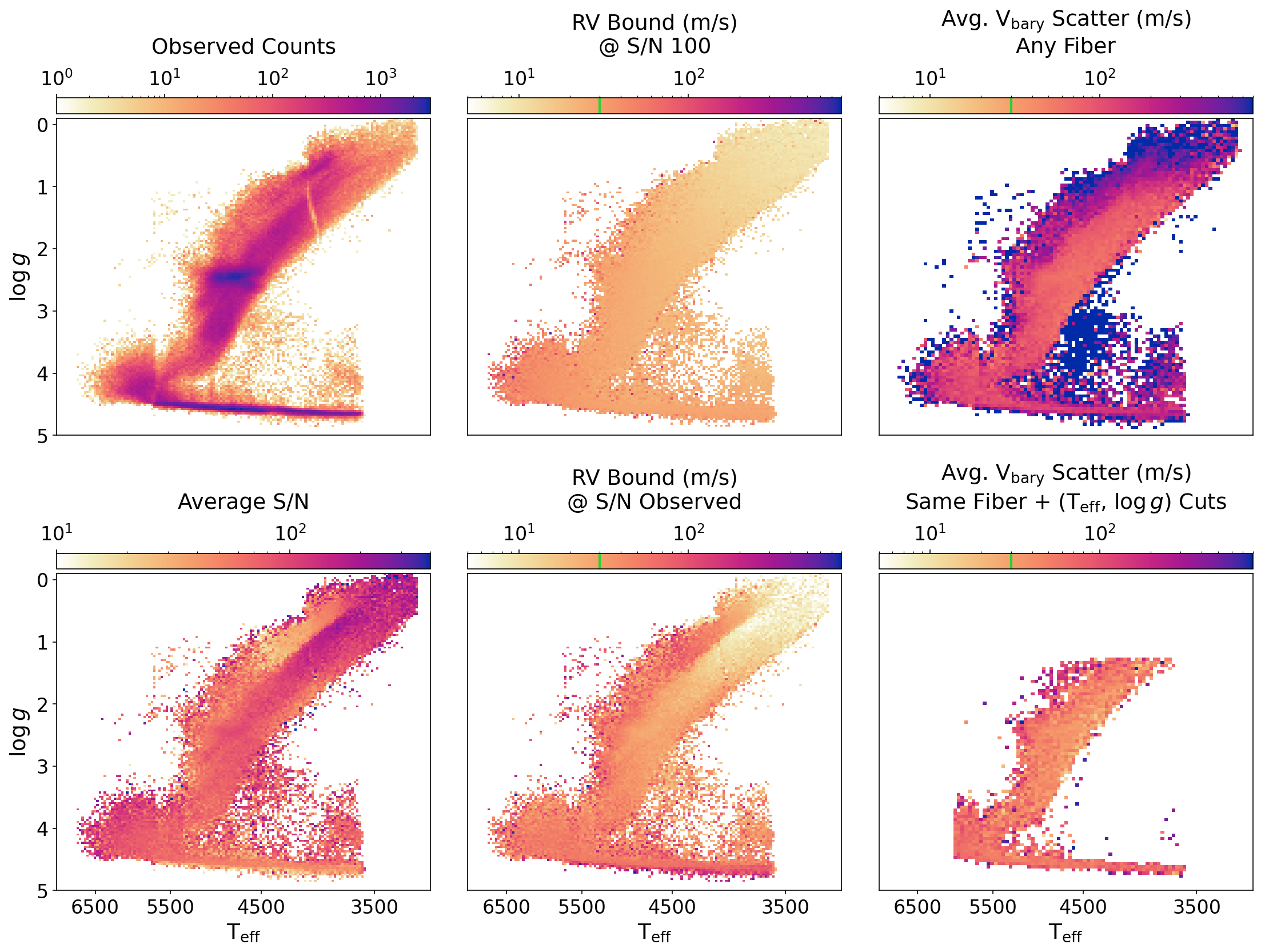}
\caption{Top Left: Logarithmic histogram of visits in APOGEE DR17 passing quality cuts versus estimates of $T_{\rm{eff}}$ and $\log g$ from ASPCAP. Bottom Left: Median S/N per visit. Middle Top: Theoretical RV lower bound if all visits were S/N 100. Middle Bottom: Theoretical RV lower bound given observed per-pixel uncertainties. Top Right: Scatter in barycentric velocity for repeat visits, irrespective of the fiber on which a star was measured. Bottom Right: Scatter in barycentric velocity for repeat visits restricted to using the same fiber.
}
\label{fig:kielVscatter}
\end{figure*}

In the preceding sections we have ignored stellar type completely. However, stellar type has a strong influence on the radial velocity precision we can achieve. For example, it is much easier to measure small shifts in a spectrum with many narrow, large amplitude lines than one with few, broad features. \cite{Bouchy_2001_A&A} formulated the ``Q-factor'' (Equation \ref{eq:Qdef}) to express the lower bound possible on the radial velocity precision (Equation \ref{eq:Qfactor}) for a given stellar spectrum and instrument response function (line spread function, throughput, etc.). This lower bound is obtained by an optimal weighting (Equation \ref{eq:Wdef}) per pixel based on the expected change for an infinitesimal shift of a given spectrum.

\begin{ceqn}
\begin{align} \label{eq:Qfactor}
    \frac{\sigma_v}{c} &= \frac{1}{\sqrt{\sum_i W_i}} = \frac{Q^{-1}}{\sqrt{\sum_i A_i}} \approx \frac{Q^{-1}}{\sqrt{N_{\rm{pix}}}*\rm{SNR}} \\
    \label{eq:Qdef}
    Q &= \frac{\sqrt{\sum_i W_i}}{\sqrt{\sum_i A_i}} \\
    \label{eq:Wdef}
    W_i &= \frac{\lambda_i^2}{\sigma^2_i}\left(\frac{\partial A_i}{\partial \lambda}\right)^2
\end{align}
\end{ceqn}

Here $\sigma_v$ is the lower bound on the radial velocity precision, $c$ is the speed of light, $A_i$ is the value of the theoretical stellar spectrum in pixel $i$, $N_{\rm{pix}}$ is the total number of pixels, $\rm{SNR}$ is the ``average'' signal-to-noise ratio of the spectrum, $\lambda$ is wavelength, and $\sigma^2_i$ is the variance for the measurement in pixel $i$. The approximation in Equation \ref{eq:Qfactor} assumes that the wavelength dependence of the measurement uncertainties does not change between observations, is entirely from photon noise (Poisson shot noise) with uniform gain, and that the uncertainties of different observations only differ by a scalar. To be concrete, this approximation is violated by additive noise, like read-noise or dark-noise, or by variable gain across the detector(s). If these conditions hold, the approximate equality becomes exact when the ``average'' signal-to-noise ratio (S/N) used to characterize the spectrum is the root-mean-square S/N. 

To help assess our RV performance as a function of stellar type, we compute these lower bounds at two different levels of approximation. First, for all 1.6 million visits with complete stellar parameters ($T_{\rm{eff}}$, $\log g$, and [X/H] not NaN), we synthesize a theoretical spectrum with \korg{} \citep{korg1, korg2} and compute the lower bound on RV precision using the observed per-pixel uncertainties, the stellar continuum component for \apMADGICS{} (which includes effects from the wavelength dependent instrument throughput), and the LSF of the fiber on which it was observed. Because we have already synthesized the spectrum (the computationally expensive step), we also compute the ``Q-factor'' again using the LSF of the fiber on which it was measured and assuming uniform S/N to have a more general expectation for the S/N dependence of the RV precision for that star. See Appendix \ref{sec:QfactorChips} for a more detailed description of the choice of average ``S/N''.

In the top left panel of Figure \ref{fig:kielVscatter}, we show a logarithmic histogram of the visits as a function of the $\log g$ and $T_{\rm{eff}}$ estimated by ASPCAP \citep{GarciaPerez_2016_AJ} that have ($T_{\rm{eff}}$, $\log g$, and [X/H] not NaN) and are well-modeled by ASPCAP (see Section \ref{sec:Data}). The sharp quasi-linear decrease in density across the giant branch near 4000K is the result of an edge of a subgrid in ASPCAP that needs to be masked by ASPCAP quality flags. The bottom left panel shows the median S/N for stars passing the same cuts over the Kiel diagram. The sharp drop in S/N perpendicular to the giant branch is associated with the boundary between the nearer, metal rich Galactic disk and more distant, lower metallicity halo.

The results from the Q-factor computations are summarized in the middle column of Figure \ref{fig:kielVscatter}. In the top panel, we show the median RV precision lower bound across the Kiel diagram if all stars had been measured to a uniform S/N of 100. This illustrates expected trends of order $\sim20$ m/s, with better RV precision for giants and worse RV precision for hotter stars. The bottom panel shows the RV precision lower bound based on per pixel flux uncertainties actually observed. A green line on the color bar indicates 30 m/s in both plots. By comparing the top and bottom panels we see that, based on the non-uniform S/N of the DR17 observations alone, we expect strong type-dependence in the RV precision. Stars in the main sequence and giants in the halo typically have S/N $<$ 100 and giants in the disk typically have S/N $>$ 100.

\begin{figure}[b]
\centering
\includegraphics[width=\linewidth]{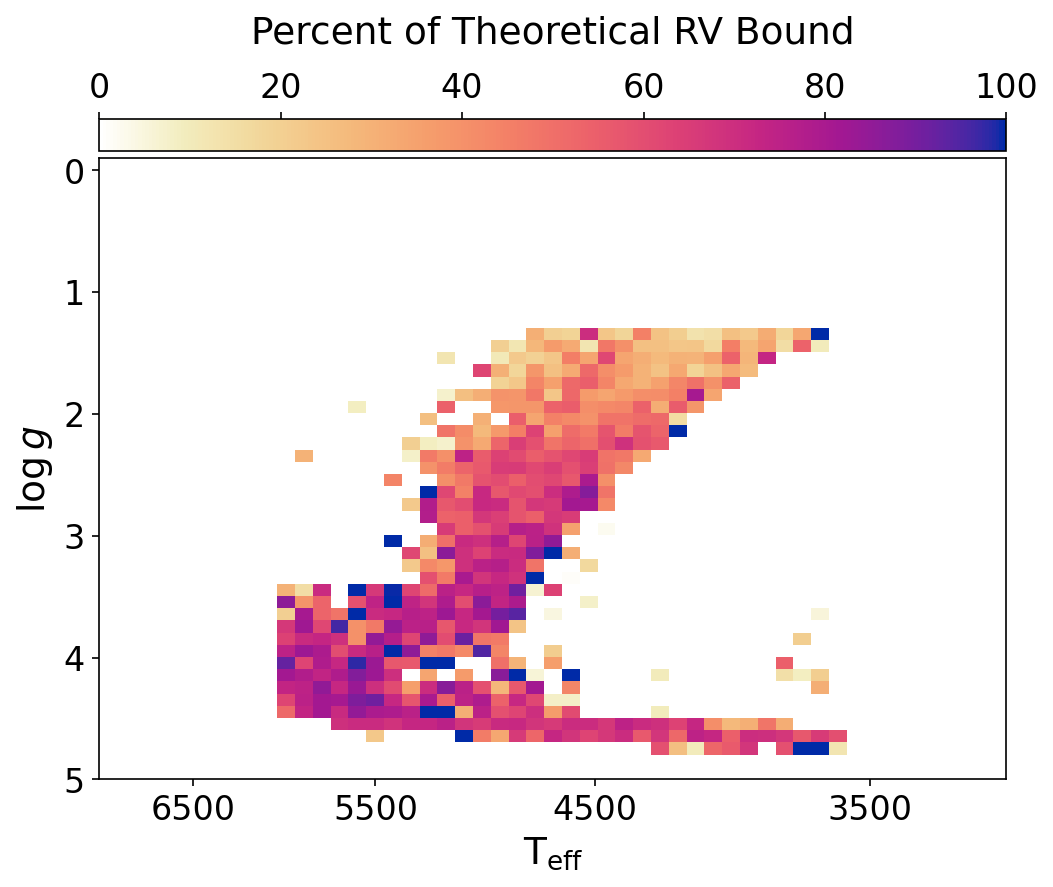}
\caption{Percentage of theoretical RV precision bound achieved by \apMADGICS{} RV catalog as measured by scatter on repeat visits to a given star using the same fiber. Obtained via a ratio of the bottom middle and bottom left plots in Figure \ref{fig:kielVscatter}.
}
\label{fig:ratioRVprec}
\end{figure}

\begin{figure*}[t]
\centering
\includegraphics[width=\linewidth]{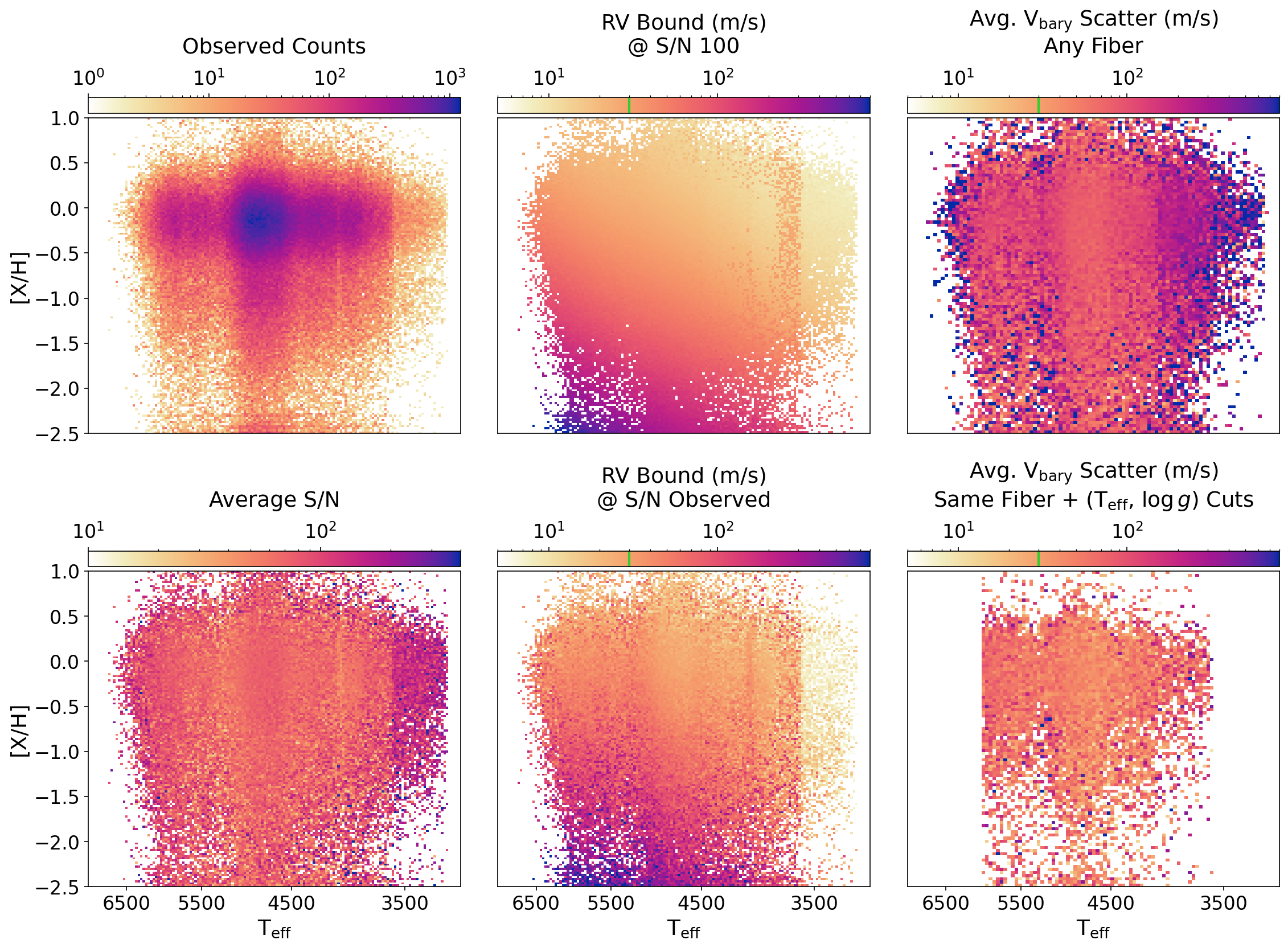}
\caption{Same as Figure \ref{fig:kielVscatter} but versus $T_{\rm{eff}}$ and [X/H].
}
\label{fig:kielMVscatter}
\end{figure*}

This provides us with a baseline expectation for a lower bound on our RV precision as a function of stellar type. In the last column of Figure \ref{fig:kielVscatter} we show the median estimate for the RV scatter from repeat visits of the same star (described in Section \ref{sec:RVPrecision}) over the Kiel diagram. In the top panel, we show this scatter, using repeat visits regardless of the fiber used to measure the spectrum. Compared to the bottom middle panel, clearly the observed RV scatter is larger than the photon-limited RV lower bound for all stellar types. This discrepancy is largest for young stellar objects and the lowest surface gravity giants, which is likely driven by a combination of flaring, turbulence, and poor stellar models, given that it is especially hard to accurately model the complex lines in cool giants. The bottom panel shows the result of applying the same cuts on $T_{\rm{eff}}$ and $\log g$ as in Figure \ref{fig:helioScatterVersusSNR} and limiting the RV scatter to repeat visits of a star using the same fiber. 

Comparing these single fiber results to the expected RV precision lower bound given the observed S/N (middle column, bottom panel), we see that for most stellar types (within these cuts), the RV precision from repeat measurements is close to the RV precision lower bound. For the highest S/N giants (S/N $\sim 100$), this translates to $\sim30$ m/s. For main sequence stars, it depends on temperature, but is closer to $\sim60$ m/s. The outliers in the bottom right panel all tend to occur in regions of stellar parameter space that are extremely sparsely populated, which may suggest an issue with the data reduction, a stellar type not well handled by ASPCAP, or a stellar type not well-captured by our low-rank stellar covariance prior. Because they represent such a small fraction of the observations, we do not treat them further.

More quantitatively, we show the ratio of Figure \ref{fig:kielVscatter} bottom middle and bottom right in Figure \ref{fig:ratioRVprec}. We find that for stars with $\log(g) > 3.5$ (main sequence stars), the \apMADGICS{} RV precision is reaching 70\% of the theoretical prediction on average, and 57\% for all stars passing our cuts. These percentages are within 10\% of the number of ``effective pixels'' without significant contributions from sky lines (see Section \ref{sec:RVPrecision}). For giants, it is also likely the the theoretical predictions overestimate the RV precision because we have not accounted for convective broadening or rotation ($v \sin(i)$).

Figure \ref{fig:kielMVscatter} is similar to Figure \ref{fig:kielVscatter}, but has the ``metallicity'' [X/H] from ASPCAP on the y-axis. The middle column, top panel of Figure \ref{fig:kielMVscatter} shows the expected diagonal trend where the RV lower bound improves with increasing metallicity and decreasing temperature. This trend is maintained for the RV lower bounds computed using the observed uncertainties. The average scatter in repeat measurements restricted to repeats using the same fiber again closely matches the theoretical bounds, but repeats pooling over different fibers do not.

When working at high precision, we always want to ask ourselves: ``Should we be doing even better?'' These analyses (Figure \ref{fig:kielVscatter} and \ref{fig:ratioRVprec}) demonstrate that with the advances implemented by \apMADGICS{} we are essentially achieving the theoretical photon-limited RV precision lower bound for repeat measurements using the same fiber. This means that, no, we cannot do better for the typical S/N currently being achieved for a given stellar type in the SDSS through DR17. However, we have shown that the LSF models fail to fully capture the fiber-to-fiber variations and mitigating these fiber-dependent variations are essential to obtaining a catalog pooled over all APOGEE visits at the instrument-limited precision.

While comforting, this means that we have not really established the systematics floor on the APOGEE instruments. The floor in Figure \ref{fig:helioScatterVersusSNR} is likely due to stellar type, at least in part, and the lack of high S/N measurements of RV stable stars. The true systematics floor of APOGEE is expected to further decrease in SDSS-V as the result of external upgrades to the instruments. First is the addition of a Fabry–Pérot interferometric calibration source, to improve the quality of the wavelength solution and LSF models, that can illuminate all 300 fibers during calibrations and 2 reference fibers during the observations. Second is a back pressure regulation system to stabilize the internal liquid nitrogen temperature, reducing distortions from expansion/contraction. And third, the addition of octagonal fiber segments near the front of the fiber train to improve radial scrambling within the fibers and achieve more uniform illumination of the detector, decoupled from fiber positioning \citep{Wilson_2022_SPIE_EXT}.

Thus, we must contend with the question ``Can we do better?'' To answer this will require the use of new data/methods for evaluating the RV precision with APOGEE. One way to further constrain the systematic floor on RV precision of APOGEE is intentionally target RV stable stars (without significant turbulence) for repeat measurements at high S/N. Another option is to emulate the exoplanet community and establish RV precision by measuring known small amplitude oscillations or the amplitude of residuals to a model fit for a time-series of RVs on a set of standard targets. This method also has the benefit of establishing the RV stability over time. In the next section, we show a test case demonstrating the benefit \apMADGICS{} brings to time series RVs for APOGEE, though in this example the signal amplitudes are large, a few km/s.

\section{RV Variability of HD 167858} \label{sec:RVVary}

\begin{figure}[t]
\centering
\includegraphics[width=\linewidth]{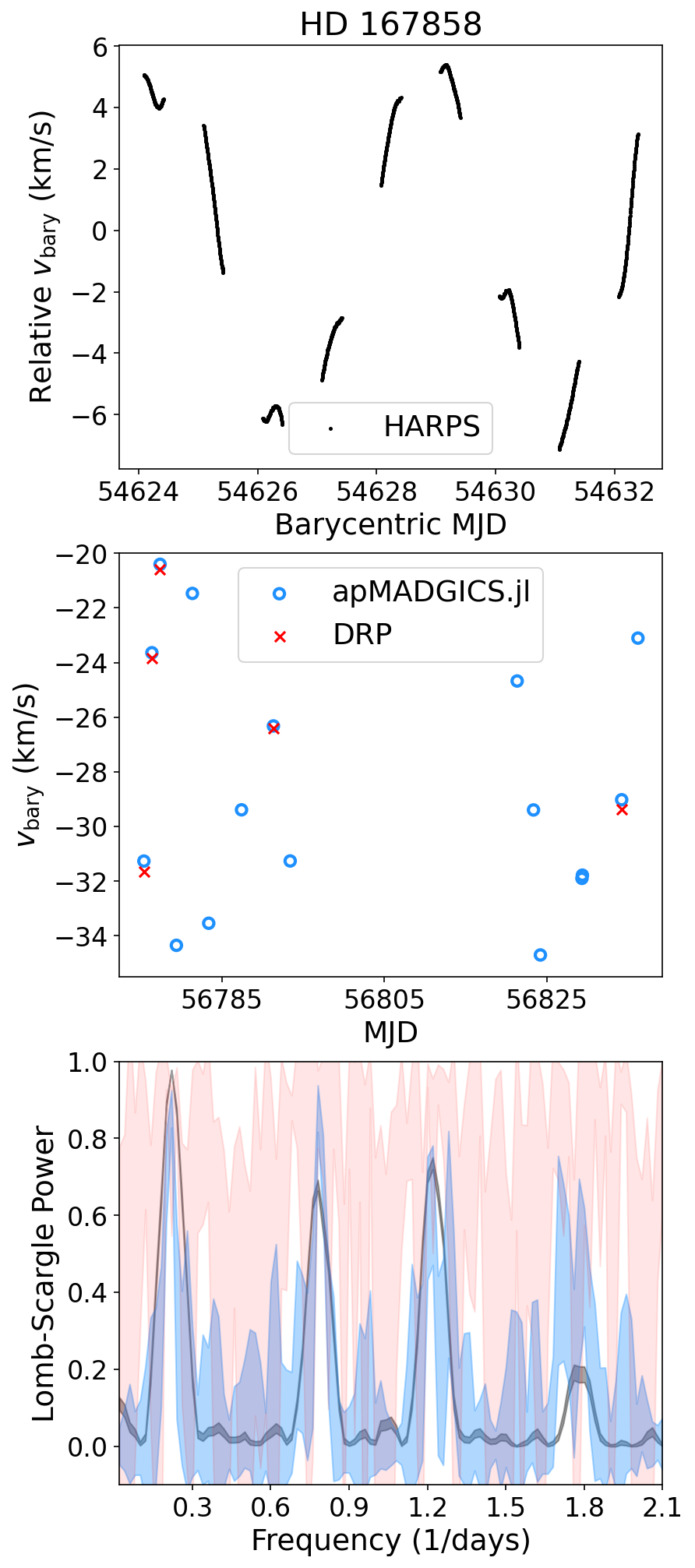}
\caption{Radial velocity (RV) observations of HD 167858 by HARPS (top) and APOGEE (middle). The RVs from both the DRP (red x) and \apMADGICS{} (blue circles) are shown for all available APOGEE visits. (Bottom) The Lomb-Scargle periodogram of these RV observations, with shading indicating the $\pm1 \sigma$ uncertainties on the periodogram, estimated by Jackknife resampling.
}
\label{fig:HARPS_LS}
\end{figure}

One focus of \apMADGICS{} is on high quality RV measurements at the visit level, treating stars as fundamentally variable objects. A test case that demonstrates the success of \apMADGICS{} at enabling variability studies is HD 167858 (Gaia DR3 ID 4275035697119244928, HR 6844), which is a candidate $\gamma\ Dor$ variable \citep{Henry_2022_AJ} targeted as a telluric standard star by APOGEE. To serve as a reference for ``ground truth,'' we use measurements of this star from the public HARPS radial velocity database \citep{Trifonov_2020_A&A}. 

The High Accuracy Radial velocity Planet Searcher (HARPS) is a $R = 120,000$, visible wavelength (378 - 691 nm), fiber-fed, echelle spectrograph on the ESO 3.6m telescope at La Silla Observatory that can achieve RV precisions of $\sim1$ m/s. Specifically, we use the differential RVs measured with SERVAL \citep{Zechmeister_2018_A&A} and corrected for systematic errors (see \citet{Trifonov_2020_A&A} for more). In general, using higher resolution spectrographs for cross-survey validation is difficult because the focus on differential radial velocity measurements means that there are often (stellar type dependent) offsets from one RV catalog to another, which limits the nature of cross-validation comparisons.

In the top panel of Figure \ref{fig:HARPS_LS}, we show 1071 RVs from HARPS (out of a total of 1084) at a typical cadence of 4 min, which samples the variability of HD 167858 in exquisite detail. In the middle panel, we show the APOGEE RVs from \apMADGICS{} with a typical cadence of 2 days. While the DRP and \apMADGICS{} RVs agree closely (up to the zeropoint offset) when both have measurements, the DRP returns no RV measurement for $\sim60\%$ of the visits. In general, across all of APOGEE DR17, \apMADGICS{} has $\sim1\%$ of its RVs with nonzero flag value while $\sim3\%$ of the RVs from the DRP are NaNs. However, understanding which RVs are trustworthy in each catalog, especially in the low S/N limit, would require a more thorough evaluation. \apMADGICS{} saves the $\Delta\chi^2$ surfaces to enable this investigation. 

In the bottom panel, we show the Lomb-Scargle periodograms over the same frequency axis, set by the APOGEE sampling and baseline, that captures the first four frequency peaks observed by the HARPS data (black line), which have corresponding amplitudes from 6 to 2 km/s. Shaded regions represent the $\pm1 \sigma$ uncertainties on the periodogram, estimated by Jackknife resampling. The \apMADGICS{} periodogram clearly detects the same four peaks, while the reduced sampling by the DRP cannot. The lowest frequency peak corresponds to the orbital frequency of HD 167858, which has a fainter binary companion (not detected in the spectra). The second frequency peak matches previously observed variability both in ground-based photometry and radial velocities ascribed to pulsation \citep{Fekel_2003_AJ}. The higher frequency peaks are likely associated with higher g-mode pulsation overtones \citep{Handler_2013_pss4}.

This case study illustrates the power of the improved visit level RVs from \apMADGICS{} and the stability of the instrument and RV measurements over month-long timescales. While the HARPS observations yield smaller uncertainties on the periodogram, Figure \ref{fig:HARPS_LS} illustrates that APOGEE was able to access many of the key science questions with $\sim 1/16^{\rm{th}}$ the total exposure time, half the collecting area, and $\sim 1/5^{\rm{th}}$ the spectral resolution. Yet, in contrast to most precision RV instruments that have a single science fiber, APOGEE can obtain quality RVs all-sky with 300 fibers in the north and south simultaneously.

\section{Data/Code Availability} \label{sec:dataavil}

The RV catalog will be more thoroughly documented and released along with all data products associated with the \apMADGICS{}\footnote{\url{https://github.com/andrew-saydjari/apMADGICS.jl}} pipeline \citep{APOGEE_pipeline_unpub}. The RV catalog, a Jupyter notebook, and data products necessary to reproduce all plots in this work are available on \href{https://doi.org/10.5281/zenodo.13251248}{Zenodo} (1.3 GB). The code used for computing Q-factors for all of APOGEE DR17 is available on Zenodo and the core functions have been integrated into \korg{}\footnote{\url{https://github.com/ajwheeler/Korg.jl}} as of v0.35.0.

\section{Conclusion} \label{sec:conc}

In this work, we present the radial velocities obtained from a re-reduction of all of APOGEE DR17 at the visit level with a new pipeline, \apMADGICS{}, based on a Bayesian component separation technique. We demonstrate that this new method, which estimates RVs while marginalizing over stellar type as well as sky line and telluric components, significantly improves the RV precision over the standard APOGEE DRP. The \apMADGICS{} RVs fully halve the systematics floor and achieve 30 m/s at high S/N for repeated measurements with the same fiber. By computing the stellar-type and S/N-dependent theoretical RV precision lower bounds, we show that \apMADGICS{} RVs are consistent to within 10\% of the theoretical sky and photon-noise limited precision possible with our instrument and current observations.

We demonstrate a degradation in the RV precision of the catalog to only $\sim56$ m/s when combined over visits observed with different fibers. We find fiber-dependent offsets in RVs that suggest this degradation is the result of unmodeled fiber-to-fiber LSF variation and identify precision LSF modeling as a key area of improvement unique to APOGEE's high precision and multiplexing. We derive a fiber-dependent calibration to mitigate this issue and achieve $\sim47$ m/s after calibration. We ensure well-calibrated RV uncertainties across all S/N for this fiber-unified catalog by leveraging repeat visits to ``RV-stable'' targets, improving on the previously underestimated RV uncertainties from the DRP. The absolute RV zeropoint is tied to our stellar models and differs from the DRP by only 100 m/s. 

In a case study, we show that \apMADGICS{} increases APOGEE's potential for multiepoch RV work and stably delivers RVs over at least month-long timescales. Similar extended multiepoch observations or higher S/N measurements of ``RV-stable'' standards will be necessary to fully validate the effect of several exciting instrument improvements made to APOGEE in SDSS-V with respect to increasing RV precision. Carefully crafting interfiber repeat observations of several standards in SDSS-V would also help further constrain the calibration of RV offsets per fiber.

\pagebreak
\section*{Acknowledgments} \label{sec:ack}
A.K.S. acknowledges support by a National Science Foundation Graduate Research Fellowship (DGE-1745303). Support for this work was provided by NASA through the NASA Hubble Fellowship grant HST-HF2-51564.001-A awarded by the Space Telescope Science Institute, which is operated by the Association of Universities for Research in Astronomy, Inc., for NASA, under contract NAS5-26555. D.P.F. acknowledges support by NSF grant AST-1614941, “Exploring the Galaxy: 3-Dimensional Structure and Stellar Streams.” D.P.F. acknowledges support by NASA ADAP grant 80NSSC21K0634 “Knitting Together the Milky Way: An Integrated Model of the Galaxy’s Stars, Gas, and Dust.” We acknowledge helpful discussions with Eddie Schlafly, David Nidever, Nathan De Lee, David Charbonneau, and Sam Quinn.

We acknowledge the support by Center for High-Performance Computing staff Brian Haymore, Anita Orendt, and Martin Cuma at the University of Utah. This work also used the Cannon cluster supported by the FAS Division of Science Research Computing Group at Harvard University. 

This work was supported by the National Science Foundation under Cooperative Agreement PHY-2019786 (The NSF AI Institute for Artificial Intelligence and Fundamental Interactions). The authors acknowledge Interstellar Institute's program ``With Two Eyes'' and the Paris-Saclay University's Institut Pascal for hosting discussions that nourished the development of the ideas behind this work.

Funding for the Sloan Digital Sky Survey V has been provided by the Alfred P. Sloan Foundation, the Heising-Simons Foundation, the National Science Foundation, and the Participating Institutions. SDSS acknowledges support and resources from the Center for High-Performance Computing at the University of Utah. SDSS telescopes are located at Apache Point Observatory, funded by the Astrophysical Research Consortium and operated by New Mexico State University, and at Las Campanas Observatory, operated by the Carnegie Institution for Science. The SDSS web site is \url{www.sdss.org}.

SDSS is managed by the Astrophysical Research Consortium for the Participating Institutions of the SDSS Collaboration, including the Carnegie Institution for Science, Chilean National Time Allocation Committee (CNTAC) ratified researchers, Caltech, the Gotham Participation Group, Harvard University, Heidelberg University, The Flatiron Institute, The Johns Hopkins University, L'Ecole polytechnique f\'{e}d\'{e}rale de Lausanne (EPFL), Leibniz-Institut f\"{u}r Astrophysik Potsdam (AIP), Max-Planck-Institut f\"{u}r Astronomie (MPIA Heidelberg), Max-Planck-Institut f\"{u}r Extraterrestrische Physik (MPE), Nanjing University, National Astronomical Observatories of China (NAOC), New Mexico State University, The Ohio State University, Pennsylvania State University, Smithsonian Astrophysical Observatory, Space Telescope Science Institute (STScI), the Stellar Astrophysics Participation Group, Universidad Nacional Aut\'{o}noma de M\'{e}xico, University of Arizona, University of Colorado Boulder, University of Illinois at Urbana-Champaign, University of Toronto, University of Utah, University of Virginia, Yale University, and Yunnan University.

\facility{Sloan/APO, Sloan/LCO}

\software{Julia \citep{bezanson2017julia},
FITSIO.jl \citep{Pence_2010_A_A},
HDF5.jl \citep{hdf5},
}

\begin{figure*}[t]
    \centering
    \includegraphics[width=\linewidth]{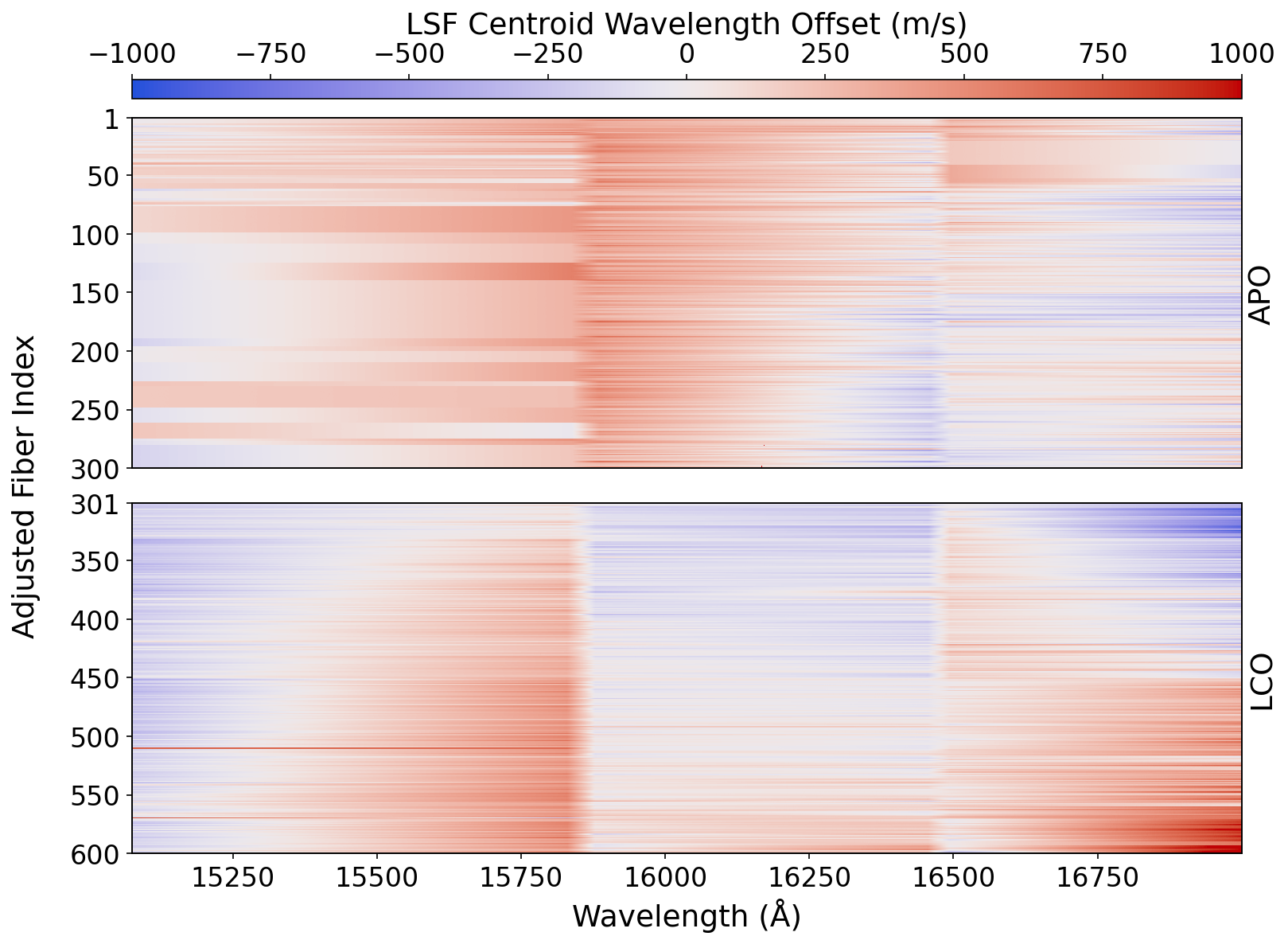}
    \caption{The offset in wavelength between the center of a Gaussian fit to the LSF model and the nominal center wavelength of the LSF model from the DRP, reported as a velocity shift in meters per second.
    }
    \label{fig:LSFCentroid}
\end{figure*}

\appendix
\begin{appendices}


\section{LSF Centroiding} \label{sec:LSFCentroid}

The wavelength solutions are derived from a least-squares fit of a Gaussian to reference lines, even though the LSF is non-Gaussian. In this section we assess how this discrepancy could impact RV precision. We queried the LSF model from the APOGEE DRP at each wavelength in the standard log-wavelength spaced grid on which we report APOGEE spectra (8700 bins) as a function of the extremely high resolution of the synthetic spectral models we generate (0.01 \r{A}) from \korg{} (200,001 bins). At each wavelength, we perform a least-squares fit of a Gaussian to the LSF. We can convert the difference between the measured central wavelength of the LSF model and the wavelength at which the LSF was queried to a velocity shift, which we show in Figure \ref{fig:LSFCentroid}. 

First, there appears to be uneven regularization in the DRP which most strikingly fixes the properties of the LSF for adjacent fiber indices on the blue chip for APO, but does so almost no where else. We also observed three fibers (510, 511, 570) at LCO (LCO FIBERID 91, 90, 31) which have large offsets on the blue chip but appear decently behaved for the longer wavelength chips. Fiber 570 is one of the ``uncalibratable'' fibers due to the lack of good RV visits and fibers 510 and 511 had the largest amplitude offsets in the fiber-fiber RV calibration (Section \ref{sec:RV_rel_fib_cal}). All of the fibers are at the edge of a v-groove block in the instrument; These edge fibers are susceptible to damage during installation, resulting in lower throughput. At both telescopes, there are broad spatially coherent trends in the centroid offsets, with the gradients at LCO being strongest.

Because the LSFs of APOGEE are \emph{not} Gaussian, the evolution of the center of a Gaussian fit likely reflects, at least in part, real structure related to the non-Gaussianity of the LSF. The point of Figure \ref{fig:LSFCentroid} is to show that the effect of this non-Gaussianity is large relative to the RV precision we are achieving, suggesting it must be handled with high accuracy. The outliers, rough regularization, and observation that we \emph{do} see fiber-dependent offsets in the RVs suggests that improving the LSF model is a key avenue for development in SDSS-V. 

These results also have important implications for the wavelength calibration of APOGEE, which fits its wavelength solution to the centers of Gaussian fits of reference lines. In DR17, these were isolated lines in ThArNe and UNe hollow-cathode lamp exposures and sky lines. In the future, in SDSS-V, these lines will also include more densely and uniformly spaced lines in Fabry-Pérot interferometer exposures. If all of the structure in Figure \ref{fig:LSFCentroid} is real non-Gaussianity, this illustrates that the Gaussian approximation will incur fiber-dependent and wavelength-dependent errors that are large relative to our RV precision. 

One approach to help ``solve'' the fiber dependence of RVs is to calibrate an RV zeropoint per fiber per night by fitting the solar spectrum to twilight spectra, assuming the wavelength solution distortions appear at leading order as a simple, type-independent offset. However, the wavelength solution distortions will still make sky line and telluric model correction more difficult. These issues of fiber-fiber consistency at the $\sim$ 30 m/s level are new and challenging exactly because of the unique positioning of APOGEE on the frontier of multiplexed precision RV work. We look forward to exploring and addressing these issues in future work.

\section{Q-factors: Chips and Fibers} \label{sec:QfactorChips}

It is convenient to have a definition of S/N for a given spectrum that acts as a single number summary. In this section, we explore what the best approximation to S/N is with respect to describing the RV precision of the observation.

\begin{figure}[t]
\centering
\includegraphics[width=0.8\linewidth]{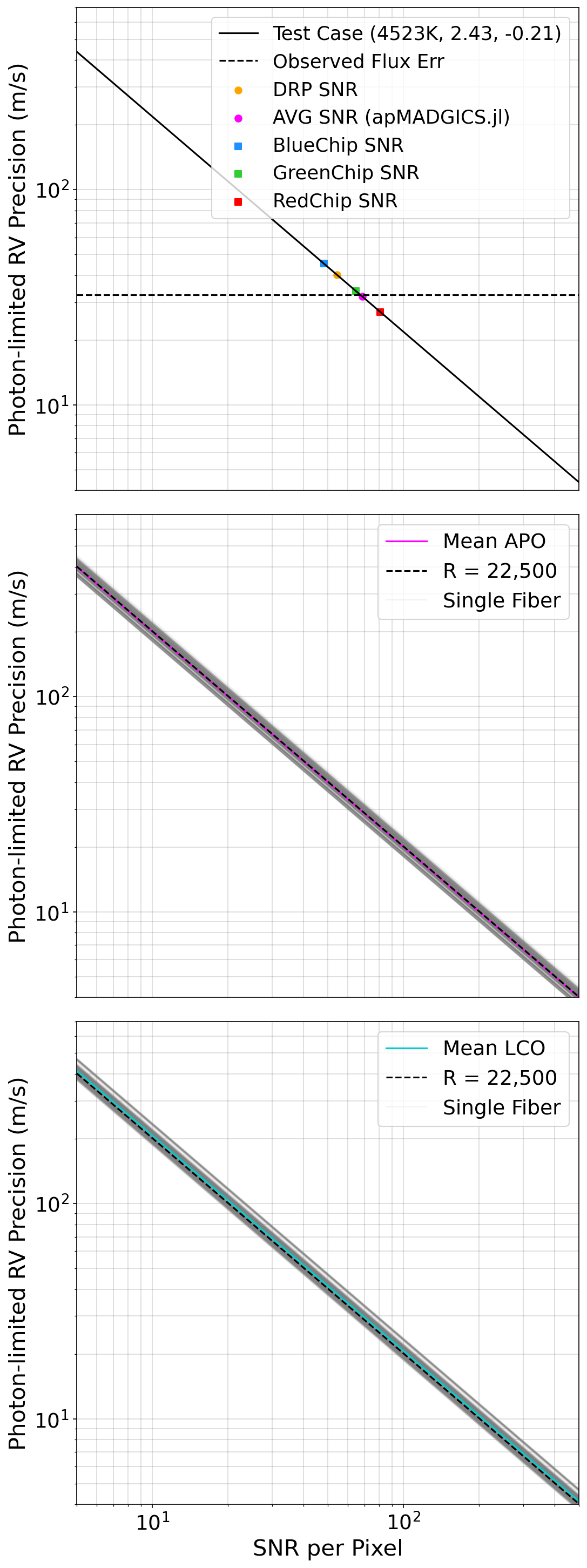}
\caption{Top panel shows how different estimates of S/N compare to the RV precision predicted using the observed flux uncertainties. Bottom two panels show fiber-to-fiber variation in predicted RV precision at APO (middle) and LCO (bottom), respectively. See text for more.
}
\label{fig:QSNRFiber}
\end{figure}

For an arbitrary test spectrum ($T_{\rm{eff}}$, $\log g$, [X/H]) = (4523 K, 2.43, -0.21) observed with fiber 17 (APO FIBERID 284), we show in Figure \ref{fig:QSNRFiber} (top panel) the photon-limited lower bound on the RV precision as a function of S/N per pixel expected from the Q-factor calculation described in Section \ref{sec:RVType} (solid line). We also show the single value for the RV precision lower bound calculated using the observed per pixel uncertainties as a horizontal dashed line. As far as discussing RV precision, the intersection of these lines identifies the ``best'' single S/N to use to describe the spectrum. If the observation was photon shot noise dominated with uniform gain, the root-mean-square S/N over the spectrum would be exactly this intersection. However, many instrumental effects (e.g. read noise, varying gain) break this equality.

To evaluate different approximate S/N definitions, we can measure how close they come to the horizontal dashed line. In Figure \ref{fig:QSNRFiber} we compare the median per pixel S/N per detector (chip), the S/N reported by the DRP, and the median S/N per pixel across the full spectrum, the last being the definition of the S/N used by \apMADGICS{}. In this test case, it is clear that the \apMADGICS{} S/N is closest to the intersection point. Comparison over all of DR17 showed that the DRP S/N estimate was systematically biased low (predicts worse RV precision) compared to the RV precision lower bound calculated using the observed per pixel uncertainties, while the difference to estimates using the median S/N per pixel across the full spectrum were symmetric.

In the middle and bottom panels of Figure \ref{fig:QSNRFiber}, we show (partially transparent grey) the RV precision lower bound for this test case if it had been observed with each of the 300 fibers at APO (middle) or LCO (bottom). The width of the shaded grey region shows the variability in the RV precision expected based solely on the fiber with which a star is observed. The mean across all fibers (for this test spectrum) is shown in solid color in both subplots and a true Gaussian LSF with R = 22,500 is shown as a dashed line for reference. The mean fiber behavior, with respect to RV precision, is almost exactly the often quoted R = 22,500 approximation for APOGEE, with LCO having only slightly lower RV precision on average. Further, the observed variation in the RV precision fiber-to-fiber is overall small.

\section{Computational Considerations} \label{sec:CompSpeed}

The run computing Q-factors for all 1.6 million visits with complete stellar parameters ($T_{\rm{eff}}$, $\log g$, and [X/H] not NaN) was carried out on 6 SDSS-owned nodes running Rocky Linux 8 (4.18.0-447.15.1) on the Center for High-Performance Computing (CHPC) cluster at the University of Utah. The nodes were equipped with AMD EPYC 7702P processors, which support AVX2 instructions and have 64 physical cores (128 hyperthreads, 16 cores/socket), 205 GB/s bandwidth per socket (12.8 GB/s/core), 500 GB total (DDR4) RAM (shared across 4 NUMA nodes), 16 MB L3 caches shared by 4 cores (256 MB total), and per core 512 kB L2 and 32 kB L1 caches. The runs used Julia v1.10.0 (single-threaded) and \korg{} v0.30.0 and took 15.4k core-h total (34.8 core-s/star), where the runtime was dominated by spectral synthesis using \korg{}. In our implementation of Equation \ref{eq:Qfactor}, we take the derivative of the spectrum with respect to wavelength by computing the difference between neighboring pixels at the extremely high resolution (0.01 \r{A}) of the synthesized model and then convolving with the LSF.

\end{appendices}

\bibliography{apMADGICS.bib}{}
\bibliographystyle{aasjournal}
\end{document}